\documentclass[preprintnumbers,
amsmath,
prd,nofootinbib,floatfix,11pt,
]{revtex4}

\newcommand\beq{\begin{eqnarray}}
\newcommand\eeq{\end{eqnarray}}
\def\dR{N_c}

\def\topoA{C}
\def\topoB{D}
\def\topoX{E}
\def\topoY{F}
\def\masterJ{J}
\def\masterK{K}
\def\masterL{L}
\def\masterP{P}
\def\masterQ{Q}
\def\masterR{R}
\def\masterS{S}
\def\lnbar{\overline{\ln}}
\def\lsim{\mathrel{\rlap{\lower4pt\hbox{$\sim$}}
    \raise1pt\hbox{$<$}}}                
\def\gsim{\mathrel{\rlap{\lower4pt\hbox{$\sim$}}
    \raise1pt\hbox{$>$}}}            

\allowdisplaybreaks
\def\MSbar{\overline{\rm MS}}
\def\metric{g}
\usepackage{setspace}
\usepackage{axodraw}
\begin{document}
\renewcommand{\theequation}{\arabic{section}.\arabic{equation}}
\renewcommand{\thefigure}{\arabic{section}.\arabic{figure}}
\renewcommand{\thetable}{\arabic{section}.\arabic{table}}

\preprint{NSF-KITP-13-248}

\title{\large \baselineskip=16pt 
Three-loop Standard Model effective potential at leading order\\ 
in strong and top Yukawa couplings}

\author{Stephen P.~Martin}
\affiliation{
\mbox{\it Department of Physics, Northern Illinois University, DeKalb IL 60115}, \\
\mbox{\it Fermi National Accelerator Laboratory, P.O. Box 500, Batavia IL 60510}, {\em and} \\
\mbox{\it Kavli Institute for Theoretical Physics,University of California, Santa Barbara, CA 9310}
}

\begin{abstract}\normalsize \baselineskip=14pt 
I find the three-loop contribution to the effective potential for the 
Standard Model Higgs field, in the approximation that the strong and top 
Yukawa couplings are large compared to all other couplings, using 
dimensional regularization with modified minimal subtraction. Checks 
follow from gauge invariance and renormalization group invariance. I 
also briefly comment on the special problems posed by Goldstone boson 
contributions to the effective potential, and on the numerical impact of 
the result on the relations between the Higgs vacuum expectation value, 
mass, and self-interaction coupling.
\end{abstract}

\maketitle
\tableofcontents
\baselineskip=15.4pt
\setcounter{footnote}{1}
\setcounter{figure}{0}
\setcounter{table}{0}

\newpage

\section{Introduction\label{sec:intro}}
\setcounter{equation}{0}
\setcounter{figure}{0}
\setcounter{table}{0}
\setcounter{footnote}{1}

The discovery \cite{ATLASHiggs,CMSHiggs,ATLAScombination,CMScombination} 
of the Standard Model Higgs boson with a mass near 126 GeV, and so far 
no other new fundamental physics, implies that a new era of precision 
analyses of the minimal electroweak symmetry breaking dynamics has begun. 
The relation between the Higgs field vacuum expectation value (VEV) and 
the underlying Lagrangian parameters, as well as the issue of vacuum 
stability, can be analyzed precisely using the effective potential 
approach \cite{Coleman:1973jx,Jackiw:1974cv,Sher:1988mj}. At present, 
the Standard Model Higgs effective potential has been evaluated at 
two-loop order in \cite{Ford:1992pn}. (The extension to more general 
models, including supersymmetric ones, is given in 
\cite{Martin:2001vx}.) An intriguing aspect of the observed Higgs mass 
is that the resulting potential is in the metastable region
near the critical value associated with a very small Higgs-self 
interaction at very high energy scales. 
Analyses of the vacuum 
stability condition before the Higgs discovery were given in 
refs.~\cite{Sher:1988mj,Lindner:1988ww,Arnold:1991cv,Ford:1992mv,Casas:1994qy,Espinosa:1995se,Casas:1996aq,Isidori:2001bm,Espinosa:2007qp,ArkaniHamed:2008ym,Bezrukov:2009db,Ellis:2009tp}, 
and some of the more detailed analyses since after the Higgs mass became 
known are given in 
\cite{EliasMiro:2011aa,Alekhin:2012py,Bezrukov:2012sa,Degrassi:2012ry,Buttazzo:2013uya,Jegerlehner:2013dpa,Bednyakov:2013cpa}.

The purpose of this paper is to find the three-loop contributions to the 
effective potential of the Standard Model, in the approximation that the 
QCD coupling $g_3$ and the top-quark Yukawa coupling $y_t$ are large 
compared to the Higgs self-interaction $\lambda$ and the
electroweak gauge couplings $g$ and $g'$ and the other 
quark and lepton Yukawa couplings. In this approximation, the three-loop part 
of the effective potential is proportional to $m_t^4$, multiplied by 
terms $g_3^4$, $g_3^2 y_t^2$, and $y_t^4$, and up to cubic logarithms. 
Here $m_t$ is the field-dependent tree-level top quark mass. The 
effective potential $V_{\rm eff}(\phi)$ is found as the sum of 
one-particle-irreducible vacuum Feynman diagrams, using couplings and 
masses obtained in the presence of a classical background field $\phi$ 
whose value at the minimum of the effective potential 
coincides with the Higgs VEV. The 
effective potential will be calculated in dimensional regularization 
\cite{Bollini:1972ui,Ashmore:1972uj,Cicuta:1972jf,tHooft:1972fi,tHooft:1973mm} 
with modified minimal subtraction \cite{Bardeen:1978yd,Braaten:1981dv}. 
The result may be used to improve the accuracy and/or theoretical error 
estimates for analyses of vacuum stability and Lagrangian parameter 
determination for the Standard Model.

To establish conventions for the present paper, consider the Higgs Lagrangian
\beq
{\cal L} = 
-\partial^\mu \Phi^\dagger \partial_\mu \Phi 
-\Lambda -m^2 \Phi^\dagger \Phi
-\lambda (\Phi^\dagger \Phi)^2 ,
\eeq
where $m^2$ is the (negative) Higgs squared mass parameter, 
and $\lambda$ is the self-coupling 
in the normalization to be used in this paper, and I use the 
metric with signature ($-$$+$$+$$+$). 
The field-independent vacuum energy term $\Lambda$ 
must be included 
in order to maintain renormalization scale invariance of the full effective 
potential and a proper treatment of renormalization group 
improvement 
\cite{Yamagishi:1981qq,Einhorn:1982pp,Kastening:1991gv,Bando:1992np},\cite{Ford:1992mv}, 
but will play no direct role in the present paper.
The complex Higgs doublet field is written
\beq
\Phi(x) = \begin{pmatrix}
\frac{1}{\sqrt{2}} [\phi + H(x) + i G^0(x)]
\\
G^+(x)
\end{pmatrix},
\eeq
where $\phi$ is the real background field, and 
$H$ is the real Higgs quantum field, 
while $G^0$ and $G^+ = G^{-*}$ 
are the real neutral and complex charged Goldstone boson fields.
The effective potential is then a function of $\phi$, with a minimum that 
equals the vacuum expectation value.

The rest of this paper is organized as follows. Section 
\ref{sec:integrals} reviews the integrals that are necessary for the 
calculation. Section \ref{sec:bare} calculates the three-loop effective 
potential in terms of bare quantities, with individual diagram 
contributions provided in an Appendix. Section \ref{sec:renormalized} 
performs the re-expression of the effective potential in terms of 
renormalized quantities, to obtain the form that can be used for 
phenomenological analyses. Section \ref{sec:goldstone} comments briefly on the
special problems posed by Goldstone boson contributions to the effective potential, and
section \ref{sec:numerical} briefly discusses 
the numerical impact of the three-loop effective potential on the 
relations between the Higgs VEV, mass, and self-interaction coupling.

\section{The necessary integrals\label{sec:integrals}}
\setcounter{equation}{0}
\setcounter{figure}{0}
\setcounter{table}{0}
\setcounter{footnote}{1}

In this section, I review the results for Feynman integrals that are necessary
for the calculations in the rest of the paper. 
Euclidean momentum integrals in
\beq
d = 4 - 2 \epsilon
\eeq
dimensions are written using the notation 
\beq
\int_p &\equiv &\int \frac{d^dp}{(2\pi)^d} .
\eeq

Consider first the integrals that depend only on one squared mass scale,
to be denoted $x$ below. In this paper, $x$ will be the 
(bare) field-dependent squared mass of the top quark.
The one-loop scalar vacuum master integral is 
\beq
A &\equiv& \int_p \frac{1}{p^2 + x} \>=\> 
\frac{\Gamma(1-d/2)}{(4 \pi)^{d/2}} x^{d/2-1} 
.
\eeq
For the two-loop integrals with one mass scale that 
are relevant below, there are two master integrals:
\beq
A^2 &=& \int_p\int_q \frac{1}{(p^2+x) (q^2+x)},
\\
B &\equiv& 
x \int_p\int_q \frac{1}{(p^2+x) q^2 (p+q)^2} =  
\frac{\Gamma(3-d) \Gamma(d/2)}{\Gamma(2-d/2)} A^2.
\eeq
Here, the term ``master integral" is taken to mean one of 
the minimal set of integrals at a given loop order 
to which all others can be reduced, with coefficients 
that are ratios of polynomials in $d$, by 
elementary algebra or integration by parts \cite{IBP} 
identities. Thus, $B$ and $A^2$ are considered
distinct two-loop one-scale master integrals by this criterion, but 
\beq
\int_p\int_q \frac{1}{(p^2+x) (q^2+x) (p+q)^2} = 
\frac{2-d}{2 (d-3)}\frac{A^2}{x} 
\eeq
is not a master integral.

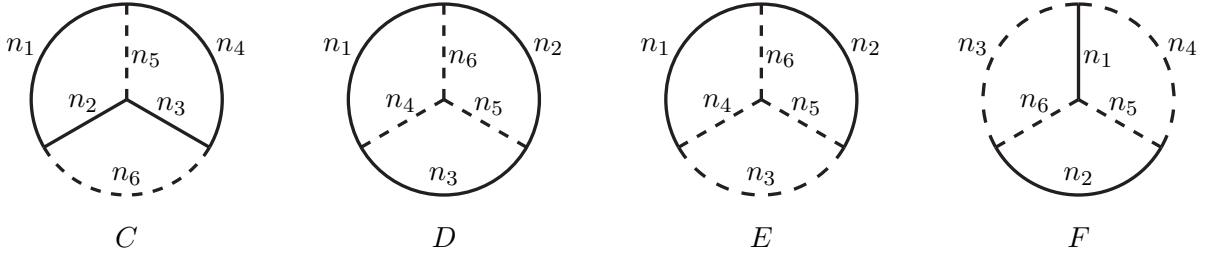
\begin{figure}[t]
\begin{center}
\SetWidth{1.25}
\begin{picture}(80,80)(-40,-40)
\CArc(0,0)(36,-30,210)
\DashCArc(0,0)(36,210,-30){4}
\DashLine(0,36)(0,0){4}
\Line(-31.1769,-18)(0,0)
\Line(31.1769,-18)(0,0)
\Text(-40,20)[]{$n_1$}
\Text(-16.5,-1.5)[]{$n_2$}
\Text(17,-2.5)[]{$n_3$}
\Text(39.5,20)[]{$n_4$}
\Text(7.4,15.5)[]{$n_5$}
\Text(0,-29)[]{$n_6$}
\Text(0,-52)[]{$\topoA$}
\end{picture}
\hspace{1.15cm}
\begin{picture}(80,80)(-40,-40)
\CArc(0,0)(36,0,360)
\DashLine(0,36)(0,0){4}
\DashLine(-31.1769,-18)(0,0){4}
\DashLine(31.1769,-18)(0,0){4}
\Text(-40,20)[]{$n_1$}
\Text(-16.5,-1.5)[]{$n_4$}
\Text(17,-2.5)[]{$n_5$}
\Text(39.5,20)[]{$n_2$}
\Text(7.4,15.5)[]{$n_6$}
\Text(0,-29)[]{$n_3$}
\Text(0,-52)[]{$\topoB$}
\end{picture}
\hspace{1.15cm}
\begin{picture}(80,80)(-40,-40)
\CArc(0,0)(36,-30,210)
\DashCArc(0,0)(36,210,330){5}
\DashLine(0,36)(0,0){4}
\DashLine(-31.1769,-18)(0,0){4}
\DashLine(31.1769,-18)(0,0){4}
\Text(-40,20)[]{$n_1$}
\Text(-16.5,-1.5)[]{$n_4$}
\Text(17,-2.5)[]{$n_5$}
\Text(39.5,20)[]{$n_2$}
\Text(7.4,15.5)[]{$n_6$}
\Text(0,-29)[]{$n_3$}
\Text(0,-52)[]{$\topoX$}
\end{picture}
\hspace{1.15cm}
\begin{picture}(80,80)(-40,-40)
\DashCArc(0,0)(36,90,210){5}
\DashCArc(0,0)(36,-30,90){5}
\CArc(0,0)(36,210,330)
\Line(0,36)(0,0)
\DashLine(-31.1769,-18)(0,0){4}
\DashLine(31.1769,-18)(0,0){4}
\Text(-40,20)[]{$n_3$}
\Text(-16.5,-1.5)[]{$n_6$}
\Text(17,-2.5)[]{$n_5$}
\Text(39.5,20)[]{$n_4$}
\Text(7.4,15.5)[]{$n_1$}
\Text(0,-29)[]{$n_2$}
\Text(0,-52)[]{$\topoY$}
\end{picture}
\end{center}
\caption{\label{fig:integrals}
From left to right, the topologies for the Feynman integrals 
$\topoA$ 
and
$\topoB$ 
and
$\topoX$ 
and
$\topoY$  defined in 
eqs.~(\ref{eq:defA}) 
and (\ref{eq:defB})
and (\ref{eq:defX})
and (\ref{eq:defY}). Solid lines 
represent propagators with squared mass $x$,
and dashed line are for massless propagators. The integers $n_i$ 
represent the powers to which
the propagators are raised, and can be positive, zero, or negative.}
\end{figure}

The three-loop one-scale Feynman integrals
encountered below can be reduced by elementary algebra 
to integrals of the types
\beq
&&\topoA(n_1,n_2,n_3,n_4,n_5,n_6) = 
\nonumber
\\
&&\qquad\int_p\int_q\int_k
\frac{1}{[(p-k)^2 + x]^{n_1} 
[p^2 + x]^{n_2} [q^2 + x]^{n_3} [(q+k)^2 + x]^{n_4} [(p+q)^2]^{n_5} [k^2]^{n_6}},\phantom{xxxxxx}
\label{eq:defA}
\\
&&\topoB(n_1,n_2,n_3,n_4,n_5,n_6) = 
\nonumber
\\
&&\qquad\int_p\int_q\int_k
\frac{1}{[(p-k)^2 + x]^{n_1}  [(q+k)^2 + x]^{n_2} [k^2 + x]^{n_3}
[p^2]^{n_4} [q^2]^{n_5} [(p+q)^2]^{n_6}},
\label{eq:defB}
\\
&&\topoX(n_1,n_2,n_3,n_4,n_5,n_6) = 
\nonumber
\\
&&\qquad\int_p\int_q\int_k
\frac{1}{[(p-k)^2 + x]^{n_1}  [(q+k)^2 + x]^{n_2} [k^2]^{n_3}
[p^2]^{n_4} [q^2]^{n_5} [(p+q)^2]^{n_6}},
\label{eq:defX}
\\
&&\topoY(n_1,n_2,n_3,n_4,n_5,n_6) = 
\nonumber
\\
&&\qquad\int_p\int_q\int_k
\frac{1}{
[(p+q)^2+x]^{n_1}
[k^2+x]^{n_2}
[(p-k)^2]^{n_3} 
[(q+k)^2]^{n_4} 
[q^2]^{n_5} 
[p^2]^{n_6} 
}
,
\label{eq:defY}
\eeq
illustrated in Figure \ref{fig:integrals}, and studied in 
\cite{Broadhurst:1991fi,Avdeev:1995eu,Broadhurst:1998rz,Schroder:2005va}. 
Here the exponents $n_i$ are integers, which can be positive, negative, or zero.
Some of the integrals with some non-positive exponents $n_i$ vanish trivially due to the
dimensional regularization identity $\int_p 1/(p^2)^n = 0$. 
Expressions involving the remaining integrals can then be systematically
simplified using the following
identities. First, re-labeling the momenta gives the 
symmetry identities:
\beq
&&
\topoA(n_1,n_2,n_3,n_4,n_5,n_6) =
\topoA(n_2,n_3,n_4,n_1,n_6,n_5) =
\topoA(n_4,n_3,n_2,n_1,n_5,n_6) ,
\label{eq:symA}
\\
&&\topoB(n_1,n_2,n_3,n_4,n_5,n_6) =
\topoB(n_1,n_3,n_2,n_6,n_5,n_4) =
\topoB(n_2,n_3,n_1,n_6,n_4,n_5) ,
\label{eq:symB}
\\
&&\topoX(n_1,n_2,n_3,n_4,n_5,n_6) =
\topoX(n_2,n_1,n_3,n_5,n_4,n_6) ,
\\
&&\topoY(n_1,n_2,n_3,n_4,n_5,n_6) =
\topoY(n_1,n_2,n_6,n_5,n_4,n_3) =
\topoY(n_2,n_1,n_5,n_4,n_3,n_6) ,\phantom{xxxxxx}
\eeq
and others obtained by repeated application of those.
Also, dimensional analysis yields:
\beq
0 &=& \Bigl [ 3d/2 - \sum_{j=1}^6 n_j + x \sum_{j=1}^4 n_j {\bf j}^+
\Bigr ] \topoA(n_1,n_2,n_3,n_4,n_5,n_6),
\label{eq:dimC}
\\
0 &=& \Bigl [ 3d/2 - \sum_{j=1}^6 n_j + x \sum_{j=1}^3 n_j {\bf j}^+
\Bigr ] \topoB(n_1,n_2,n_3,n_4,n_5,n_6),
\\
0 &=& \Bigl [ 3d/2 - \sum_{j=1}^6 n_j + x \sum_{j=1}^2 n_j {\bf j}^+
\Bigr ] \topoX(n_1,n_2,n_3,n_4,n_5,n_6),
\\
0 &=& \Bigl [ 3d/2 - \sum_{j=1}^6 n_j + x \sum_{j=1}^2 n_j {\bf j}^+
\Bigr ] \topoY(n_1,n_2,n_3,n_4,n_5,n_6).
\eeq
where here and below 
the notation for bold-faced raising and lowering operators is the standard one such that, for each integer $j=1,\ldots,6$, we have
${\bf j}^\pm \topoA(\ldots,n_j,\ldots) \equiv \topoA(\ldots,n_j\pm1,\ldots)$
and similarly for $\topoB$, $\topoX$, and $\topoY$.
Finally, integration by parts \cite{IBP} gives the identities:
\beq
0 &=& \bigl [d - n_3 - n_4 - 2 n_5
+ n_3 {\bf 3}^+({\bf 2}^- - {\bf 5}^-)
+ n_4 {\bf 4}^+({\bf 1}^- - {\bf 5}^-)
\bigr ] \topoA
,\phantom{xxx}
\label{eq:IBPtopoA}
\\
0 &=& \bigl [d - n_1 - 2 n_2 - n_5
+ n_1 {\bf 1}^+({\bf 6}^- - {\bf 2}^- + 2 x)
+ 2 x n_2 {\bf 2^+} 
+ n_5 {\bf 5}^+ ({\bf 3}^- - {\bf 2}^-)
\bigr ] \topoA
,\phantom{xxx}
\eeq
and
\beq
0 &=& \bigl [
d-n_3 - n_4 - 2 n_5
+ n_3 {\bf 3}^+ ({\bf 2}^- - {\bf 5}^-)
+ n_4 {\bf 4}^+ ({\bf 6}^- - {\bf 5}^-)
\bigr ] \topoB
,\phantom{xxx}
\\
0 &=& \bigl [
d - n_1 - 2 n_2  - n_3
+ n_1 {\bf 1}^+ ({\bf 6}^-- {\bf 2}^- +  2x )
+ 2 x n_2 {\bf 2}^+ 
+ n_3 {\bf 3}^+ ({\bf 5}^- - {\bf 2}^- +  2x )
\bigr ] \topoB
,\phantom{xxxx}
\\
0 &=& \bigl [
d - 2 n_2 - n_5  - n_6 
+ 2 x n_2 {\bf 2}^+
+ n_5 {\bf 5}^+ ({\bf 3}^- - {\bf 2}^- )
+ n_6 {\bf 6}^+ ({\bf 1}^- - {\bf 2}^- ) 
\bigr ] \topoB
,\phantom{xxxx}
\eeq
and
\beq
0 &=& \bigl [
d-n_2 - n_5 - 2 n_6
+ n_2 {\bf 2}^+ ({\bf 1}^- - {\bf 6}^-)
+ n_5 {\bf 5}^+ ({\bf 4}^- - {\bf 6}^-)
\bigr ] \topoX
,\phantom{xxx}
\\
0 &=& \bigl [
d - 2 n_3 - n_4 - n_5
+ n_4 {\bf 4}^+ ({\bf 1}^- - {\bf 3}^- - x)
+ n_5 {\bf 5}^+ ({\bf 2}^- - {\bf 3}^- - x)
\bigr ] \topoX
,\phantom{xxx}
\\
0 &=& \bigl [
d - 2 n_2 - n_5 - n_6 
+ 2 x n_2 {\bf 2}^+
+ n_5 {\bf 5}^+ ({\bf 3}^- - {\bf 2}^- + x)
+ n_6 {\bf 6}^+ ({\bf 1}^- - {\bf 2}^-)
\bigr ] \topoX
,
\\
0 &=& \bigl [
d - n_3 - 2 n_4 - n_5 
+ n_3 {\bf 3}^+ ({\bf 1}^- - {\bf 4}^- - x)
+ n_5 {\bf 5}^+ ({\bf 6}^- - {\bf 4}^-)
\bigr ] \topoX
,
\\
0 &=& \bigl [
d - n_1 - n_2 - 2n_3 
+ n_1 {\bf 1}^+ ({\bf 4}^- - {\bf 3}^- + x)
+ n_2 {\bf 2}^+ ({\bf 5}^- - {\bf 3}^- + x)
\bigr ] \topoX
,
\\
0 &=& \bigl [
d - n_1 - 2 n_2 - n_3 
+ n_1 {\bf 1}^+ ({\bf 6}^- - {\bf 2}^- + 2 x)
+ 2x n_2 {\bf 2}^+
+ n_3 {\bf 3}^+ ({\bf 5}^- - {\bf 2}^- + x)
\bigr ] \topoX
,\phantom{xxxx}
\\
0 &=& \bigl [
d - n_1 - 2 n_4 - n_6 
+ n_1 {\bf 1}^+ ({\bf 3}^- - {\bf 4}^- + x)
+ n_6 {\bf 6}^+ ({\bf 5}^- - {\bf 4}^-)
\bigr ] \topoX
,
\eeq
and
\beq
0 &=& \bigl [
d-2 n_2 - n_5 - n_6 
+ 2 x n_2 {\bf 2}^+ 
+ n_5 {\bf 5}^+ ({\bf 4}^- - {\bf 2}^- + x)
+ n_6 {\bf 6}^+ ({\bf 3}^- - {\bf 2}^- + x)
\bigr ] \topoY
,\phantom{xxx}
\\
0 &=& \bigl [
d - n_2 - 2 n_5  - n_6
+ n_2 {\bf 2}^+ ({\bf 4}^- - {\bf 5}^- + x)
+ n_6 {\bf 6}^+ ({\bf 1}^- - {\bf 5}^- - x)
\bigr ] 
\topoY
.\phantom{xxxx}
\label{eq:IBPtopoY}
\eeq
In equations (\ref{eq:IBPtopoA})-(\ref{eq:IBPtopoY}), the arguments $(n_1, n_2, n_3, n_4, n_5, n_6)$
are implicit for $\topoA$, $\topoB$, $\topoX$, and $\topoY$, and were omitted for the sake of simplicity.

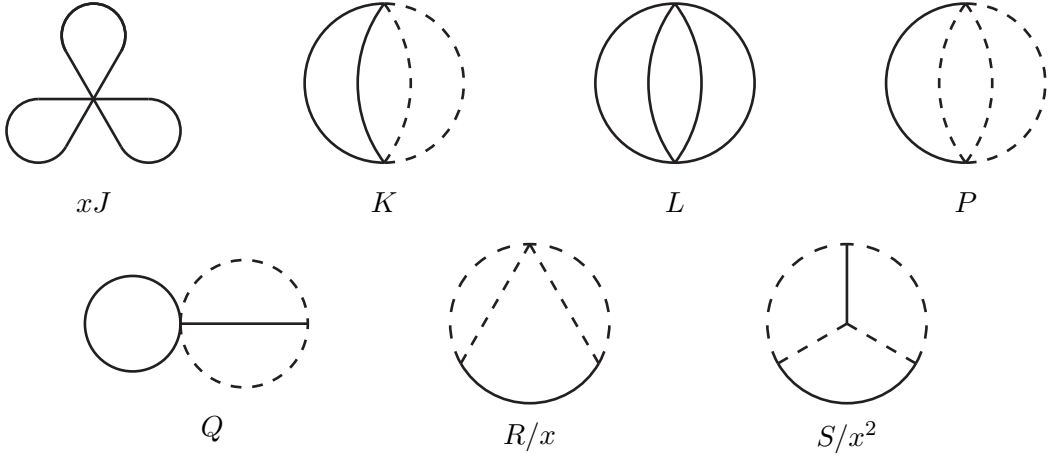
\begin{figure}[t]
\begin{center}
\SetWidth{1}
\begin{picture}(60,60)(-30,-24)
\CArc(0,24)(12,-30,210)
\CArc(20.7846,-12)(12,-150,90)
\CArc(-20.7846,-12)(12,90,330)
\CArc(0,24)(12,-30,210)
\Line(0,0)(10.3923,18)
\Line(0,0)(-10.3923,18)
\Line(0,0)(10.3923,-18)
\Line(0,0)(-10.3923,-18)
\Line(0,0)(20.7846,0)
\Line(0,0)(-20.7846,0)
\Text(0,-39)[]{$x\masterJ$}
\end{picture}
\hspace{1.5cm}
\begin{picture}(60,60)(-30,-30)
\DashCArc(0,0)(30,-90,90){4}
\CArc(0,0)(30,90,270)
\CArc(40,0)(50,143.13,216.87)
\DashCArc(-40,0)(50,-36.8699,36.8699){4}
\Text(0,-45)[]{$\masterK$}
\end{picture}
\hspace{1.5cm}
\begin{picture}(60,60)(-30,-30)
\CArc(0,0)(30,0,360)
\CArc(40,0)(50,143.13,216.87)
\CArc(-40,0)(50,-36.8699,36.8699)
\Text(0,-45)[]{$\masterL$}
\end{picture}
\hspace{1.5cm}
\begin{picture}(60,60)(-30,-30)
\DashCArc(0,0)(30,-90,90){4}
\CArc(0,0)(30,90,270)
\DashCArc(40,0)(50,143.13,216.87){4}
\DashCArc(-40,0)(50,-36.8699,36.8699){4}
\Text(0,-45)[]{$\masterP$}
\end{picture}

\begin{picture}(80,100)(-40,-40)
\DashCArc(12,0)(24,-180,180){4}
\CArc(-30,0)(18,0,360)
\Line(-12,0)(36,0)
\Text(0,-40)[]{$\masterQ$}
\end{picture}
\hspace{1.15cm}
\begin{picture}(80,100)(-40,-40)
\DashCArc(0,0)(30,-30,210){5}
\CArc(0,0)(30,210,330)
\DashLine(-25.9807,-15)(0,30){4}
\DashLine(25.9807,-15)(0,30){4}
\Text(0,-43)[]{$\masterR /x$}
\end{picture}
\hspace{1.15cm}
\begin{picture}(80,100)(-40,-40)
\DashCArc(0,0)(30,-30,210){5}
\CArc(0,0)(30,210,330)
\Line(0,30)(0,0)
\DashLine(-25.9807,-15)(0,0){4}
\DashLine(25.9807,-15)(0,0){4}
\Text(0,-43)[]{$\masterS /x^2$}
\end{picture}
\end{center}
\caption{\label{fig:masters} 
The three-loop master integrals $\masterJ$, $\masterK$, $\masterL$,
$\masterP$, $\masterQ$, $\masterR$, and $\masterS$
defined in eqs.~(\ref{eq:defmasterJ})-(\ref{eq:defmasterS}). 
Solid lines represent scalar propagators with squared mass 
$x$, and dashed line are for massless propagators.
All of the integrals in Figure \ref{fig:integrals} can be 
reduced to linear combinations of these, with coefficients that are ratios of
polynomials in the number of spacetime dimensions, $d$.}
\end{figure}

Repeated applications of the identities in equations (\ref{eq:symA})-(\ref{eq:IBPtopoY})
allows \cite{Broadhurst:1991fi,Avdeev:1995eu} 
all of the one-scale integrals of the types 
$\topoA$, $\topoB$, $\topoX$, and $\topoY$ 
to be reduced (with many redundant checks) to just the seven 
master integrals depicted in Figure \ref{fig:masters}:
\beq
J &\equiv& \topoA(0,1,1,1,0,0)/x = \topoB(1,1,1,0,0,0)/x = A^3/x,
\label{eq:defmasterJ}
\\
K &\equiv& \topoA(0,1,0,1,1,1) = \topoB(0,1,1,1,0,1) = \topoX(1,1,0,1,1,0) = \topoY(1,1,0,1,0,1),
\\
L &\equiv& \topoA(1,1,1,1,0,0),
\\
P &\equiv& \topoX(0,1,1,1,0,1),
\\
Q &\equiv& \topoX(1,1,1,0,1,0) = \topoY (1,1,0,1,1,0) = AB/x,
\\
R &\equiv& x \topoY(0,1,1,1,1,1),
\\
S &\equiv& x^2 \topoY(1,1,1,1,1,1).
\label{eq:defmasterS}
\eeq
The expressions for $\masterJ$, $\masterK$, $\masterP$, $\masterQ$, and 
$\masterR$ are known exactly in terms of gamma functions, and for the 
remaining integrals $\masterL$ and $\masterS$ the results are known as 
expansions in $\epsilon$. Actually, the integral $\masterS$ will not be 
needed in the present paper. The integrals $\masterJ$, $\masterL$, and 
$\masterP$ are needed here to order $\epsilon^2$, while $\masterK$ and 
$\masterQ$ are needed to order $\epsilon^1$, and $\masterR$ to order 
$\epsilon^0$. Writing each of the other master integrals in terms of 
$A^3/x$, the expansions to these orders (and beyond) are found from 
refs.~\cite{Broadhurst:1991fi,Broadhurst:1998rz}:
\beq
\masterJ &=& A^3/x ,
\\
\masterK &=& \frac{A^3}{x} 
\Bigl (
-\frac{1}{3} - \frac{\epsilon}{6} + \frac{5\epsilon^2}{12} +
\left [ \frac{79}{24} - \frac{8\zeta(3)}{3} \right ] \epsilon^3 
+ \left [\frac{685}{48} + \frac{2 \pi^4}{15} - \frac{4 \zeta(3)}{3} 
\right] \epsilon^4  
+ \ldots
\Bigr ) ,
\\
\masterL &=& \frac{A^3}{x} \Bigl (
-2 -\frac{5}{3} \epsilon - \frac{1}{2} \epsilon^2 + \frac{103}{12} \epsilon^3 +
  [1141/24 - 112 \zeta(3)/3] \epsilon^4 +
  \bigl [ 9055/48 + 136\pi^4/45 
\nonumber \\ &&
  + 32 \ln^2(2) [\pi^2 - \ln^2(2)]/3 
  -168 \zeta(3) - 256 {\rm Li}_4(1/2)\bigr ] \epsilon^5 + \ldots
\Bigr ) ,
\\
\masterP &=& \frac{A^3}{x} \Bigl (
\frac{\epsilon}{12} + \frac{3\epsilon^2}{8} + 
\left [67/48 + \pi^2/12 \right ] \epsilon^3
+ \left [ 457/96 + 3 \pi^2/8 - 5 \zeta(3)/6 \right ] \epsilon^4
\nonumber \\ &&
+ \left [ 2971/192 + 67 \pi^2/48 + 23 \pi^4/360 - 15 \zeta(3)/4 \right ] \epsilon^5 
+ \ldots
\Bigr ) ,
\\
\masterQ &=& \frac{A^3}{x} \Bigl (
-\frac{1}{2} - \frac{\epsilon}{2} + 
\Bigl [-1-\frac{\pi^2}{6} \Bigr ] \epsilon^2
+ \Bigl [\zeta(3) -2 - \frac{\pi^2}{6} \Bigr ] \epsilon^3
+ \Bigl [ \zeta(3) -4 - \frac{\pi^2}{3} - \frac{\pi^4}{20} \Bigr ] \epsilon^4
+ \ldots
\Bigr ) ,\phantom{xxx}
\\
\masterR &=& \frac{A^3}{x} \Bigl (
\frac{1}{3} + 
\frac{2\epsilon}{3} + [5/3 + \pi^2/3] \epsilon^2 + 
\left [4 + 2 \pi^2/3 - 4 \zeta(3)/3 \right ] \epsilon^3
+ \ldots
\Bigr ) .
\eeq
A particularly useful and 
systematic compendium of these and many other one-scale 
multi-loop vacuum integral results
can be found in ref.~\cite{Schroder:2005va}.

In the following, we will also need certain integrals that depend 
on two squared mass scales: $x$ (which, as above, will be the top-quark
squared mass) and $y$ (which will be either
the Higgs or Goldstone bare squared mass).
Because of the approximation used in this paper, it is sufficient to have
these integrals to first order in $y$, and to order $\epsilon^1$ for two-loop
integrals and $\epsilon^0$ for three-loop integrals. The two-scale 
integrals needed are:
\beq
I_{xxy} 
&\equiv& 
\int_p\int_q \frac{1}{(p^2+x) (q^2 +x) [(p+q)^2 + y]} ,
\label{eq:defIxxy}
\\
I_{x0y} 
&\equiv& 
\int_p\int_q \frac{1}{(p^2+x) q^2 [(p+q)^2 + y]} ,
\\
I_{xxxxy} &\equiv& \int_p\int_q\int_k
\frac{1}{[(p-k)^2 + x] [(q+k)^2 + x] 
[p^2 + x] [q^2 + x] [k^2+y]},\phantom{xxxxxx}
\\
I_{xx00y} &\equiv& \int_p\int_q\int_k
\frac{1}{[(p-k)^2 + x]  [(q+k)^2 + x] p^2 q^2 [k^2+y] } .
\eeq
Using integration by parts and dimensional analysis, these integrals
are found to obey the differential equations:
\beq
y (4x-y) \frac{d}{dy} I_{xxy} &=& 
(d-3)(2x-y) I_{xxy} + (d-2) \left [ 1 - (y/x)^{d/2-1} \right ] A^2 ,
\\
(x-y)^2  \frac{d}{dy} I_{x0y} &=&
(3-d)(x-y) I_{x0y} + (1-d/2)(1-y/x) (y/x)^{d/2-1} A^2 ,
\\
y (4x-y) \frac{d}{dy} I_{xxxxy} &=&
(2dx-8x+5y-3dy/2) I_{xxxxy} + (3d/2-4) L + (4-2d) A I_{xxy} ,
\phantom{xxx}
\\
2 y (y-x) \frac{d}{dy} I_{xx00y} &=&
(d x - 2 x + 3 d y - 10y) I_{xx00y} + (2d-4) A I_{x0y} + (8-3d) K .
\eeq
It follows that, to the order needed below:
\beq
I_{xxy}  &=& \frac{A^2}{x} \frac{2-d}{2(d-3)} \Bigl \{
1 + r \Bigl (\frac{1}{2} - \epsilon \ln(r) + \epsilon^2 \Bigl [
\frac{1}{2} \ln^2(r) + 2 \ln(r) - 4 \Bigr ]
\nonumber \\ &&
+ \epsilon^3 \Bigl [-\frac{1}{6} \ln^3(r)
- \ln^2(r) + 8 \Bigr ] + {\cal O}(\epsilon^4) \Bigr )
+ {\cal O}(r^2)
\Bigr \}
,
\\
I_{x0y}  &=& \frac{B}{x} \Bigl \{
1 + r \Bigl (1 - 2 \epsilon \ln(r) + \epsilon^2 \bigl [
\ln^2(r) + 2 \ln(r) - 2 - 2\pi^2/3 \bigr ]
\nonumber \\ &&
+ \epsilon^3 \Bigl [-\frac{1}{3} \ln^3(r)
- \ln^2(r) + (2 + 2 \pi^2/3) \ln(r) -2 + 2 \pi^2/3 + 4 \zeta(3) \Bigr ] 
+ {\cal O}(\epsilon^4) \Bigr ) 
\phantom{xxxxx}
\nonumber \\ &&
+ {\cal O}(r^2)
\Bigr \}
,
\\
I_{xxxxy}  &=& \frac{1}{4(4-d)}\left [
(3d-8)\frac{L}{x} + \frac{2(d-2)^2}{d-3} \frac{A^3}{x^2} \right ]
+ r \frac{A^3}{x^2} 
\Bigl \{
\frac{1}{3} 
+ \epsilon [2/3 - \ln(r)]
\nonumber \\ &&
+ \epsilon^2 \Bigl [ \frac{1}{2} \ln^2(r) + 2 \ln(r) - \frac{19}{3} \Bigr ]
+ \epsilon^3 \Bigl [ -\frac{1}{6} \ln^3(r) - \ln^2(r) - \ln(r) + 16 +
\frac{14}{3} \zeta(3) \Bigr ]
\nonumber \\ &&
+ {\cal O}(\epsilon^4) 
\Bigr \}
+ {\cal O}(r^2)
,
\\
I_{xx00y}  &=& \Bigl ( \frac{3d-8}{d-2}\Bigr ) \frac{K}{x} - 2 \frac{A B}{x^2} 
+ r \frac{A^3}{x^2} 
\Bigl \{
\frac{1}{3} 
+ \epsilon \Bigl [\frac{2}{3} - \ln(r) \Bigr ]
+ \epsilon^2 \Bigl [ \frac{1}{2} \ln^2(r) - \frac{\pi^2}{3} 
- \frac{1}{3} \Bigr ]
\nonumber \\ &&
+ \epsilon^3 \Bigl [ -\frac{1}{6} \ln^3(r) - \frac{1}{2} + \frac{\pi^2}{2}
+ \frac{14}{3} \zeta(3) \Bigr ]
+ {\cal O}(\epsilon^4) 
\Bigr \}
+ {\cal O}(r^2) ,
\label{eq:expIxx00y}
\eeq
where $r=y/x$. The above 
expansions for $I_{xxy}$ and $I_{x0y}$ can also be obtained
as special cases of results in \cite{Davydychev}, and the expansion for
$I_{xxxxy}$ can be obtained as a special case of eq.~(3.27) in \cite{Kalmykov:2005hb}.

\section{Effective potential in terms of bare quantities\label{sec:bare}}
\setcounter{equation}{0}
\setcounter{figure}{0}
\setcounter{table}{0}
\setcounter{footnote}{1}

Consider the effective potential written in terms of the 
bare external scalar field $\phi_B$
and the bare coupling parameters including the
Yukawa coupling $y_{tB}$, the strong coupling $g_{3B}$, and the Higgs self coupling $\lambda_B$. 
This is calculated in
$d=4 - 2\epsilon$ dimensions in terms of the bare parameters in the Lagrangian, without
including any counterterms. The conversion to $\MSbar$ parameters will be done in the next
section. The expansion in terms of the loop order $\ell$ reads 
\beq
V_{\rm eff} = \sum_{\ell=0}^\infty V^{(\ell)}_B,
\label{eq:Veffbare}
\eeq
where the tree-level potential in this expansion is 
\beq
V^{(0)}_B = \Lambda_B + \frac{m^2_B}{2} \phi_B^2 + \frac{\lambda_B}{4} \phi_B^4 .
\label{eq:V0bare}
\eeq
The bare field-dependent squared masses of the top quark, Higgs scalar $H^0$,
and the Goldstone bosons $G^0, G^\pm$ (in Landau gauge), 
are denoted by
\beq
x &=& y_{tB}^2 \phi_B^2/2,
\\
x_H &=& m^2_B + 3 \lambda_B \phi_B^2,
\\
x_G &=& m^2_B + \lambda_B \phi_B^2,
\eeq
and the integrals $A$, $B$, $\masterJ$, $\masterK$, $\masterL$,
$\masterP$, $\masterQ$, and $\masterR$
of the previous section are taken to be functions of $x$.
In order to maximize the generality of results below, and allow 
more informative checks,
they are written in terms of 
the group theory quantities
\beq
C_G &=& \dR = 3,
\label{eq:defCG}
\\
C_F &=& \frac{N_c^2-1}{2N_c} = 4/3,
\\
T_F &=& 1/2,
\\
N_q &=& 6,
\label{eq:defNq}
\eeq
where $C_G$ is the Casimir invariant and Dynkin index of the $SU(3)_c$ gauge group,
$C_F$, $T_F$, and $\dR$ are the Casimir invariant, Dynkin index, and dimension of
the fundamental representation, and $N_q$ is the number of quarks in the theory.

The well-known one-loop order top, Higgs, and Goldstone contributions are then:
\beq
V^{(1)}_B &=&  \frac{x A}{d} \left [ 
-4 \dR + (x_H/x)^{d/2} + 3 (x_G/x)^{d/2} \right ].
\label{eq:V1bare}
\eeq
Note that even though the aim of this 
paper is to neglect terms proportional to the Higgs and Goldstone masses 
in the three-loop order part of the
renormalized result, they do need to be included in the one-loop bare 
contribution. This is because
when $\lambda_B$ is expressed in terms of renormalized couplings, it includes
a counter-term proportional to $y_t^4$ with no $\lambda$.
The other one-loop contributions involving electroweak vector bosons and lighter
fermions are not written here, 
because after expressing bare quantities
in terms of renormalized quantities, they do not
affect the determination of the
three-loop contribution at leading order in the QCD and top-quark Yukawa couplings.

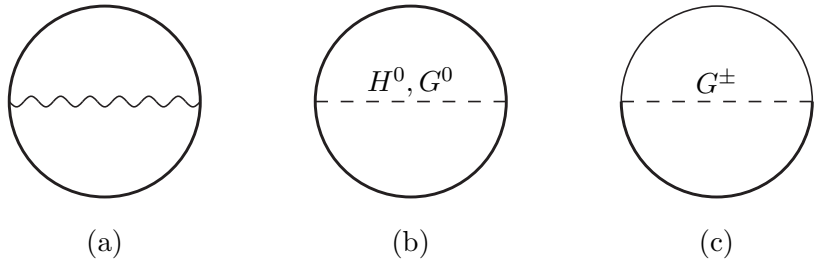
\begin{figure}[t]
\begin{center}
\begin{picture}(80,94)(-40,-54)
\SetWidth{0.6}
\Text(0,-54)[]{(a)}
\Photon(36,0)(-36,0){2}{6.5}
\SetWidth{1.1}
\CArc(0,0)(36,0,360)
\end{picture}
\hspace{1cm}
\begin{picture}(80,94)(-40,-54)
\SetWidth{0.6}
\Text(0,-54)[]{(b)}
\Text(0,7.5)[]{$H^0, G^0$}
\DashLine(36,0)(-36,0){5}
\SetWidth{1.1}
\CArc(0,0)(36,0,360)
\end{picture}
\hspace{1cm}
\begin{picture}(80,94)(-40,-54)
\SetWidth{0.6}
\CArc(0,0)(36,0,360)
\Text(0,-54)[]{(c)}
\Text(0,7.5)[]{$G^\pm$}
\DashLine(36,0)(-36,0){5}
\SetWidth{1.1}
\CArc(0,0)(36,180,360)
\end{picture}
\end{center}
\caption{\label{fig:diagrams2} The 2-loop vacuum Feynman diagrams
involving the top quark, neglecting electroweak interactions. Wavy lines are 
gluons, heavy solid lines are top quarks, lighter solid lines are bottom quarks,
and dashed lines are scalars as labeled.}
\end{figure}

At two-loop order, the pertinent contributions
are from the diagrams shown in Figure \ref{fig:diagrams2}. 
(Here, and below, each figure is taken to represent diagrams with all helicities 
and mass insertions consistent with the topology shown.)
The gluon is treated with an arbitrary
gauge-fixing parameter $\xi$, with propagator
\beq
-i [\metric^{\mu\nu}/p^2 - (1-\xi) p^\mu p^\nu/(p^2)^2],
\eeq
where $\xi=0$ for Landau gauge and $\xi=1$ for Feynman gauge. 
The dependence on the QCD $\xi$
cancels in the effective potential, providing a useful check. The combined
two-loop order result is
\beq
V^{(2)}_B &=& g_{3B}^2 \dR C_F\frac{(d-1)(d-2)}{(d-3)} A^2
+ y_{tB}^2 \dR \Bigl \{
A^2 \bigl [1 - (x_H/x)^{d/2-1} - 2 (x_G/x)^{d/2-1} \bigr ]
\nonumber \\ &&
+ (2x - x_H/2) I_{xxx_H} - x_G I_{xxx_G}/2 + (x-x_G) I_{x0x_G}
\Bigr \}
.
\label{eq:V2bare}
\eeq
Note that here one must 
include terms up to linear order in $\lambda_B$ and first order in 
$\epsilon$ from the diagrams of Figure \ref{fig:diagrams2}b,c, again
because $\lambda_B$ written in terms of renormalized couplings will contain
a term proportional to $y_t^4$ with no $\lambda$. 

The pertinent three-loop order diagrams at leading order in the strong and top-quark Yukawa couplings
are shown in
Figure \ref{fig:diagrams3}. 
All except diagrams \ref{fig:diagrams3}(r)
and \ref{fig:diagrams3}(s)
are evaluated by first writing them in terms
of the functions $\topoA$, $\topoB$, $\topoX$, and $\topoY$
defined in the previous section, and then using the identities in equations
(\ref{eq:symA})-(\ref{eq:IBPtopoY})
to reduce the result to the six master integrals 
$\masterJ$, $\masterK$, $\masterL$, $\masterP$, $\masterQ$, and
$\masterR$. 
For diagrams \ref{fig:diagrams3}(r)
and \ref{fig:diagrams3}(s), it is necessary to also make use of the two-scale
integrals in eqs.~(\ref{eq:defIxxy})-(\ref{eq:expIxx00y}), because of the 
``doubled" Higgs and Goldstone propagators.
I find:
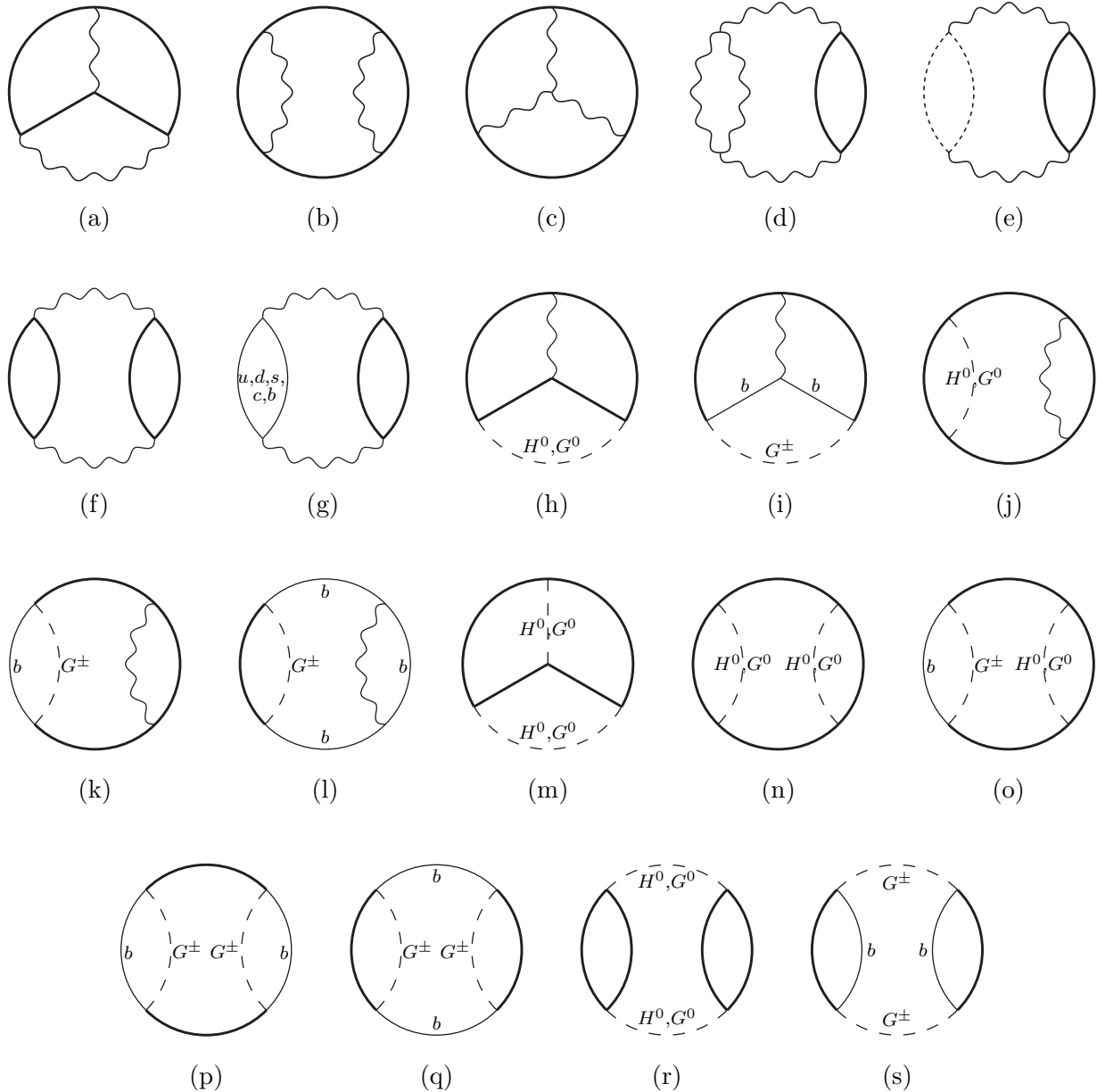
\begin{figure}[tp]
\begin{center}
\SetWidth{0.6}
\begin{picture}(80,94)(-40,-54)
\Text(0,-54)[]{(a)}
\PhotonArc(0,0)(36,210,-30){2}{5.5}
\Photon(0,36)(0,0){2}{2.5}
\SetWidth{1.1}
\Line(-31.1769,-18)(0,0)
\Line(0,0)(31.1769,-18)
\CArc(0,0)(36,-30,210)
\end{picture}
\hspace{0.365cm}
\begin{picture}(80,94)(-40,-54)
\Text(0,-54)[]{(b)}
\PhotonArc(-50.9117,0)(36,-45,45){2}{4.5}
\PhotonArc(50.9117,0)(36,135,225){2}{4.5}
\SetWidth{1.1}
\CArc(0,0)(36,0,360)
\end{picture}
\hspace{0.365cm}
\begin{picture}(80,94)(-40,-54)
\Text(0,-54)[]{(c)}
\Photon(0,36)(0,0){2}{2.5}
\Photon(-31.1769,-18)(0,0){2}{2.5}
\Photon(31.1769,-18)(0,0){2}{2.5}
\SetWidth{1.1}
\CArc(0,0)(36,0,360)
\end{picture}
\hspace{0.365cm}
\begin{picture}(80,94)(-40,-54)
\Text(0,-54)[]{(d)}
\PhotonArc(0,0)(36,45,135){2}{4.5}
\PhotonArc(0,0)(36,225,315){2}{4.5}
\PhotonArc(0,0)(36,135,225){2}{4.5}
\PhotonArc(-50.9117,0)(36,-45,45){2}{4.5}
\SetWidth{1.1}
\CArc(0,0)(36,-45,45)
\CArc(50.9117,0)(36,135,225)
\end{picture}
\hspace{0.365cm}
\begin{picture}(80,94)(-40,-54)
\Text(0,-54)[]{(e)}
\PhotonArc(0,0)(36,45,135){2}{4.5}
\PhotonArc(0,0)(36,225,315){2}{4.5}
\DashCArc(0,0)(36,135,225){1.85}
\DashCArc(-50.9117,0)(36,-45,45){1.85}
\SetWidth{1.1}
\CArc(0,0)(36,-45,45)
\CArc(50.9117,0)(36,135,225)
\end{picture}
\end{center}

\vspace{0.01cm}

\begin{center}
\begin{picture}(80,94)(-40,-54)
\Text(0,-54)[]{(f)}
\PhotonArc(0,0)(36,45,135){2}{4.5}
\PhotonArc(0,0)(36,225,315){2}{4.5}
\SetWidth{1.1}
\CArc(0,0)(36,-45,45)
\CArc(50.9117,0)(36,135,225)
\CArc(0,0)(36,135,225)
\CArc(-50.9117,0)(36,-45,45)
\end{picture}
\hspace{0.365cm}
\begin{picture}(80,94)(-40,-54)
\Text(0,-54)[]{(g)}
\Text(-26.3,0)[]{$\scriptstyle u,d,s,$}
\Text(-25,-7.5)[]{$\scriptstyle c,b$}
\PhotonArc(0,0)(36,45,135){2}{4.5}
\PhotonArc(0,0)(36,225,315){2}{4.5}
\CArc(0,0)(36,135,225)
\CArc(-50.9117,0)(36,-45,45)
\SetWidth{1.1}
\CArc(0,0)(36,-45,45)
\CArc(50.9117,0)(36,135,225)
\end{picture}
\hspace{0.365cm}
\begin{picture}(80,94)(-40,-54)
\Text(0,-54)[]{(h)}
\Text(0,-29)[]{$\scriptstyle H^0, G^0$}
\DashCArc(0,0)(36,210,-30){5}
\Photon(0,36)(0,0){2}{2.5}
\SetWidth{1.1}
\Line(-31.1769,-18)(0,0)
\Line(0,0)(31.1769,-18)
\CArc(0,0)(36,-30,210)
\end{picture}
\hspace{0.365cm}
\begin{picture}(80,94)(-40,-54)
\Text(0,-54)[]{(i)}
\Text(0,-30)[]{$\scriptstyle G^\pm$}
\Text(15,-3)[]{$\scriptstyle b$}
\Text(-15,-3)[]{$\scriptstyle b$}
\Line(-31.1769,-18)(0,0)
\Line(0,0)(31.1769,-18)
\DashCArc(0,0)(36,210,-30){5}
\Photon(0,36)(0,0){2}{2.5}
\SetWidth{1.1}
\CArc(0,0)(36,-30,210)
\end{picture}
\hspace{0.365cm}
\begin{picture}(80,94)(-40,-54)
\Text(0,-54)[]{(j)}
\Text(-15,0)[]{$\scriptstyle H^0,G^0$}
\DashCArc(-50.9117,0)(36,-45,45){5}
\PhotonArc(50.9117,0)(36,135,225){2}{4.5}
\SetWidth{1.1}
\CArc(0,0)(36,0,360)
\end{picture}
\end{center}

\vspace{0.01cm}

\begin{center}
\begin{picture}(80,94)(-40,-54)
\Text(0,-54)[]{(k)}
\Text(-8,0)[]{$\scriptstyle G^\pm$}
\Text(-33,-1)[]{$\scriptstyle b$}
\CArc(0,0)(36,0,360)
\DashCArc(-50.9117,0)(36,-45,45){5}
\PhotonArc(50.9117,0)(36,135,225){2}{4.5}
\SetWidth{1.1}
\CArc(0,0)(36,-135,135)
\end{picture}
\hspace{0.365cm}
\begin{picture}(80,94)(-40,-54)
\Text(0,-54)[]{(l)}
\Text(-8,0)[]{$\scriptstyle G^\pm$}
\Text(33,-1)[]{$\scriptstyle b$}
\Text(0,31)[]{$\scriptstyle b$}
\Text(0,-31)[]{$\scriptstyle b$}
\CArc(0,0)(36,0,360)
\DashCArc(-50.9117,0)(36,-45,45){5}
\PhotonArc(50.9117,0)(36,135,225){2}{4.5}
\SetWidth{1.1}
\CArc(0,0)(36,135,225)
\end{picture}
\hspace{0.25cm}
\begin{picture}(80,94)(-40,-54)
\Text(0,-54)[]{(m)}
\Text(0,-29)[]{$\scriptstyle H^0, G^0$}
\Text(0,15)[]{$\scriptstyle H^0, G^0$}
\DashCArc(0,0)(36,210,-30){5}
\DashLine(0,36)(0,0){5}
\SetWidth{1.1}
\Line(-31.1769,-18)(0,0)
\Line(0,0)(31.1769,-18)
\CArc(0,0)(36,-30,210)
\end{picture}
\hspace{0.365cm}
\begin{picture}(80,94)(-40,-54)
\Text(0,-54)[]{(n)}
\Text(-15,0)[]{$\scriptstyle H^0,G^0$}
\Text(15,0)[]{$\scriptstyle H^0,G^0$}
\DashCArc(-50.9117,0)(36,-45,45){5}
\DashCArc(50.9117,0)(36,135,225){5}
\SetWidth{1.1}
\CArc(0,0)(36,0,360)
\end{picture}
\hspace{0.365cm}
\begin{picture}(80,94)(-40,-54)
\Text(0,-54)[]{(o)}
\Text(-8,0)[]{$\scriptstyle G^\pm$}
\Text(-33,-1)[]{$\scriptstyle b$}
\Text(15,0)[]{$\scriptstyle H^0,G^0$}
\CArc(0,0)(36,0,360)
\DashCArc(-50.9117,0)(36,-45,45){5}
\DashCArc(50.9117,0)(36,135,225){5}
\SetWidth{1.1}
\CArc(0,0)(36,-135,135)
\end{picture}
\end{center}

\vspace{0.01cm}

\begin{center}
\begin{picture}(80,94)(-40,-54)
\Text(0,-54)[]{(p)}
\Text(-8,0)[]{$\scriptstyle G^\pm$}
\Text(-33,-1)[]{$\scriptstyle b$}
\Text(8,0)[]{$\scriptstyle G^\pm$}
\Text(33,-1)[]{$\scriptstyle b$}
\CArc(0,0)(36,0,360)
\DashCArc(-50.9117,0)(36,-45,45){5}
\DashCArc(50.9117,0)(36,135,225){5}
\SetWidth{1.1}
\CArc(0,0)(36,-135,-45)
\CArc(0,0)(36,45,135)
\end{picture}
\hspace{0.365cm}
\begin{picture}(80,94)(-40,-54)
\Text(0,-54)[]{(q)}
\Text(-8,0)[]{$\scriptstyle G^\pm$}
\Text(0,31.5)[]{$\scriptstyle b$}
\Text(0,-31)[]{$\scriptstyle b$}
\Text(8,0)[]{$\scriptstyle G^\pm$}
\CArc(0,0)(36,0,360)
\DashCArc(-50.9117,0)(36,-45,45){5}
\DashCArc(50.9117,0)(36,135,225){5}
\SetWidth{1.1}
\CArc(0,0)(36,-45,45)
\CArc(0,0)(36,135,225)
\end{picture}
\hspace{0.365cm}
\begin{picture}(80,94)(-40,-54)
\Text(0,-54)[]{(r)}
\Text(0,29)[]{$\scriptstyle H^0, G^0$}
\Text(0,-29)[]{$\scriptstyle H^0, G^0$}
\DashCArc(0,0)(36,45,135){5}
\DashCArc(0,0)(36,225,315){5}
\SetWidth{1.1}
\CArc(0,0)(36,-45,45)
\CArc(50.9117,0)(36,135,225)
\CArc(0,0)(36,135,225)
\CArc(-50.9117,0)(36,-45,45)
\end{picture}
\hspace{0.365cm}
\begin{picture}(80,94)(-40,-54)
\Text(0,-54)[]{(s)}
\Text(0,29)[]{$\scriptstyle G^\pm$}
\Text(0,-29)[]{$\scriptstyle G^\pm$}
\CArc(50.9117,0)(36,135,225)
\CArc(-50.9117,0)(36,-45,45)
\Text(11,0)[]{$\scriptstyle b$}
\Text(-11,0)[]{$\scriptstyle b$}
\DashCArc(0,0)(36,45,135){5}
\DashCArc(0,0)(36,225,315){5}
\SetWidth{1.1}
\CArc(0,0)(36,-45,45)
\CArc(0,0)(36,135,225)
\end{picture}
\end{center}

\caption{\label{fig:diagrams3} The 3-loop vacuum Feynman diagrams
involving the top quark and strong interactions. Wavy lines are 
gluons, heavy solid lines are top quarks, lighter solid lines are
other quarks as labeled, dashed lines are $H^0$, $G^0$ and $G^\pm$ as labeled,
and dotted lines are ghosts.}
\end{figure}
\beq
V^{(3)}_B &=& g_{3B}^4 \dR C_F \Bigl \{ C_G  \Bigl [
  \frac{(2-d)^3}{2 (d-4)^2 (d-3)} \masterJ
  + \frac{(d-2)^2 (2 d^2 - 17 d + 32)}{2 (d-4) (2d-7)} \masterK
\nonumber \\ &&
  + \frac{(3-d)(d^3 - 13 d^2 + 50 d - 48)}{4 (d-4)^2} \masterL
\Bigr]
\nonumber \\ &&
+ C_F \Bigl [
\frac{(d-2)^2 (-d^5 + 13 d^4 - 67 d^3 + 181 d^2 - 274 d + 188)}{
2 (d-4)^2 (d-3)^2} \masterJ
\nonumber \\ &&
+
\frac{(2-d)(2 d^3 - 21 d^2 + 67 d - 68)}{(d-4)(d-3)} \masterK
+
\frac{(d-6)(d-3)(d^2 - 7d + 8)}{2(d-4)^2} \masterL
\Bigr ]
\nonumber \\ &&
+ T_F \Bigl [
\frac{2(5-d)(d-2)^3}{(d-6)(d-4)(d-3)} \masterJ
+ \frac{d^3 - 7 d^2 + 6 d + 16}{(d-6)(4-d)} \masterL 
+ (N_q-1) 
\frac{4 (d-3)(d-2)}{7 - 2 d} \masterK
\Bigr ]
\Bigr \} 
\nonumber \\ &&
+ g_{3B}^2 y_{tB}^2 \dR C_F \Bigl \{
\frac{2 (2-d) (2 d^3 - 17 d^2 + 48 d - 46)}{(d-4)^2 (d-3)^2} \masterJ
+ \frac{2 (2-d) (2d-5)}{(d-4)(d-3)} \masterK
\nonumber \\ &&
+ \frac{(3-d)(d^2 + 2d - 16)}{2(d-4)^2} \masterL
+ \frac{(d-2)(d^3 - d^2 - 16 d + 32)}{2 (d-4)^2 (d-3)} \masterP 
\nonumber \\ &&
+ \frac{(2-d)(2 d^3 - 11 d^2 + 9 d + 16)}{2 (d-4)(d-3)} \masterQ
\Bigr \}
\nonumber \\ &&
+y_{tB}^4 \dR \Bigl \{
\frac{-d^4 + 13 d^3 - 52 d^2 + 68 d -8}{4 (d-4)^2 (d-3)^2} \masterJ
+
\frac{3 d^4 - 34 d^3 + 182 d^2 - 480 d + 480}{2 d (d-2)(d-3)(d-4)} \masterK
\nonumber \\ &&
+ \frac{(6-d)(3d-8)}{8 (d-4)^2} \masterL 
+ \frac{4 d^2 - 21 d + 28}{2 (d-4)(d-3)} \masterP
+ \frac{3 (2-d)}{8} \masterR
+ \frac{2 d^2 -d - 12}{2 d (d-3)} \masterQ
\Bigr \}
\nonumber \\ &&
+y_{tB}^4 \dR^2 \Bigl \{
\Bigl [\frac{2 (d-2)}{(d-4)(d-3)} 
+ \frac{d-2}{2} (x_H/x)^{d/2-2}
+ \frac{3(d-2)}{4} (x_G/x)^{d/2-2}
\Bigr ] \masterJ
+ \frac{18-7d}{2 (d-2)} \masterK
\phantom{xxx}
\nonumber \\ &&
+ \frac{d-2}{d-4} \masterL
+ 3 \masterQ 
+ 4 x \frac{d}{dx_H} \Bigl [ x I_{xxxxx_H}- A I_{xxx_H} \Bigr ]
+ x \frac{d}{dx_G} \Bigl [\frac{x}{2} I_{xx00x_G}- A I_{x0x_G} \Bigr ]
\Bigr \} .
\label{eq:V3bare}
\eeq
The individual diagram contributions, exhibiting the
separate dependences on $\xi$, are shown in the Appendix. 
The task of the next section is to re-express these results in 
terms of $\MSbar$ renormalized
quantities.

\section{Effective potential in terms of renormalized 
quantities\label{sec:renormalized}}
\setcounter{equation}{0}
\setcounter{figure}{0}
\setcounter{table}{0}
\setcounter{footnote}{1}

The effective potential in the $\MSbar$ renormalization scheme is 
obtained by re-expressing the bare quantities in terms of
renormalized quantities. Write:
\beq
\phi_B &=&  \mu^{-\epsilon} \phi \sqrt{Z_\phi} ,
\label{eq:defphiB}
\\
Z_\phi &=& 
1 + \sum_{\ell=1}^\infty \sum_{n=1}^\ell 
\frac{c^\phi_{\ell,n}}{(16\pi^2)^\ell\epsilon^n} ,
\label{eq:defZphi}
\\
z_{kB} &=&  \mu^{\rho_{k} \epsilon} \Bigl ( 
z_k + \sum_{\ell=1}^\infty \sum_{n=1}^\ell 
\frac{c^k_{\ell,n}}{(16\pi^2)^\ell \epsilon^n}
\Bigr ),
\label{eq:defzkB}
\eeq
where the subscript $B$ indicates bare quantities, the absence of
a subscript $B$ indicates an $\MSbar$ renormalized quantity, 
$\ell$ is the loop order, and 
$k$ is an index that runs over the list of Lagrangian parameters, including
$z_k = \lambda, y_t, g_3, m^2, \Lambda$, with\footnote{As a 
simplifying notation, in subscripts and superscripts 
a specific parameter $z_k$ is used interchangeably with 
the corresponding index $k$, so that $\rho_k \equiv \rho_{z_k}$ and 
$c^{k}_{\ell,n} \equiv c^{z_k}_{\ell,n}$ and 
$\beta_{k} \equiv \beta_{z_k}$.} 
$\rho_\lambda = 2$, $\rho_{g_3} = \rho_{y_t} = 1$, 
$\rho_{m^2} = 0$, $\rho_\Lambda = -2$. 
The mass scale $\mu$ is the (arbitrary) dimensional regularization scale,
introduced so that $\int d^dx V$ is dimensionless,
and so that $g_3$ and $y_t$ are also dimensionless,
and the field-dependent top-quark mass $y_t\phi /\sqrt{2}$ 
has mass dimension 1, for any $\epsilon$.
The regularization scale $\mu$ is related to the $\MSbar$ renormalization scale $Q$ by
\cite{Bardeen:1978yd,Braaten:1981dv}
\beq
Q^2 = 4\pi e^{-\gamma_E} \mu^2.
\eeq
where $\gamma_E = 0.5772\ldots$ is the Euler-Mascheroni constant. The 
counter-term quantities $c^{\phi}_{\ell,n}$ and $c^k_{\ell,n}$ are 
polynomials in the $\MSbar$ renormalized parameters $z_k$, and are 
independent of $\epsilon$ and $\phi$. They are determined by the 
requirement that the full effective potential and other physical 
quantities have no poles in $\epsilon$ when expressed in terms of 
$\MSbar$ quantities.

The $\MSbar$ beta functions and the scalar anomalous dimension are defined by
\beq
\beta_{k} &\equiv&
Q \frac{d z_k}{dQ} \Bigl |_{\epsilon=0}
\>=\>
Q \frac{d z_k}{dQ} + \epsilon \rho_k z_k ,
\\
\gamma &\equiv& 
-Q \frac{d \ln\phi}{dQ} \Bigl |_{\epsilon=0}
\>=\> 
-Q \frac{d \ln\phi}{dQ} + \epsilon
\>=\> 
\frac{1}{2} Q \frac{d}{dQ} \ln(Z_\phi) 
.
\eeq
It is useful to write these as loop expansions:
\beq
\beta_{k} &=& 
\sum_{\ell=1}^\infty \frac{1}{(16 \pi^2)^\ell} \beta_{k}^{(\ell)},
\\
\gamma &=& 
\sum_{\ell=1}^\infty \frac{1}{(16 \pi^2)^\ell} \gamma^{(\ell)}.
\eeq
Now, by using the fact that the bare quantities $\phi_B$ and $z_{kB}$ 
cannot depend on $Q$ (or $\mu$), 
one obtains the beta functions and anomalous dimension
in terms of the simple pole counterterms:
\beq
\beta_{k}^{(\ell)} &=& 2 \ell c^k_{\ell,1} ,
\label{eq:betafromck}
\\
\gamma^{(\ell)} &=& -\ell c^\phi_{\ell,1},
\label{eq:gammaafromcphi}
\eeq
as well as the consistency conditions for 
higher pole counterterms with $\ell \geq n \geq 2$:
\beq
2 \ell c^k_{\ell,n}
&=&
\sum_{\ell'=1}^{\ell-n+1}\sum_j \beta_{j}^{(\ell')} 
\frac{\partial}{\partial z_j}
c^k_{\ell-\ell',n-1}
,
\label{eq:higherpoleck}
\\
\ell c^\phi_{\ell,n} &=&
\sum_{\ell'=1}^{\ell-n+1} \Bigl (
-\gamma^{(\ell')} + 
\frac{1}{2} \sum_j \beta_{j}^{(\ell')} \frac{\partial}{\partial z_j}
\Bigr ) c^\phi_{\ell-\ell',n-1}
.
\label{eq:higherpolecphi}
\eeq
The identities
\beq
\sum_{j} \rho_{j} z_j \frac{\partial }{\partial z_j} c^k_{\ell,n}
&=& (2 \ell + \rho_k) c^k_{\ell,n} ,
\\
\sum_{j} \rho_{j} z_j \frac{\partial }{\partial z_j} c^\phi_{\ell,n}
&=& 2 \ell c^\phi_{\ell,n}
\eeq
have been used to simplify the preceding expressions.
[Note that eq.~(2.11) in ref.~\cite{MVI} has a missing factor of $-\rho_k$
on the left side.] 

Equations (\ref{eq:betafromck})-(\ref{eq:higherpolecphi})
allow the coefficients $c^k_{\ell,n}$ and $c^\phi_{\ell,n}$ to be determined
from the known results for the beta functions and scalar anomalous dimension.
The ones that are needed for this paper are
\cite{MVI,MVII,Jack:1984vj,MVIII}, \cite{Ford:1992pn}, 
\cite{Chetyrkin:2012rz} 
(see also \cite{Chetyrkin:2013wya,Bednyakov:2013eba}):
\beq
c^{\lambda}_{1,1} &=& 
-\dR y_t^4   
+ \lambda y_t^2 (2 \dR)
+ 12 \lambda^2 + \ldots
,\\
c^{\lambda}_{2,1} &=& 
g_3^2 y_t^4 (-2 \dR C_F)  
+ y_t^6 (5 \dR /2)
+ \lambda g_3^2 y_t^2 (5 \dR C_F) 
+ \lambda y_t^4 (-\dR/4) 
+ \ldots
,\\
c^{\lambda}_{2,2} &=& g_3^2 y_t^4 (6 \dR C_F)  
+ y_t^6 (-2 \dR^2 - 3 \dR/2) 
+ \lambda g_3^2 y_t^2 (-6 \dR C_F)
\nonumber \\ &&
+ \lambda y_t^4 (3 \dR^2 - 21 \dR/2)
+ \ldots
\\
c^{\lambda}_{3,1} &=& g_3^4 y_t^4 \dR C_F \left \{[8 \zeta(3) -109/6] C_G 
+ [131/6 - 16 \zeta(3)] C_F + (16 + 10 N_q/3) T_F  
\right \}
\nonumber \\ &&
+ g_3^2 y_t^6 \dR C_F [20 \zeta(3) - 19/6]
+ y_t^8 \dR [-4 \zeta(3) + 13/6 - 65\dR/8] 
+ \ldots
,\\
c^{\lambda}_{3,2} &=& g_3^4 y_t^4  \dR C_F \left (
24 C_G + 10 C_F - 16 T_F N_q/3  
\right ) + g_3^2 y_t^6 \dR C_F (-25 - 9 \dR) 
\nonumber \\ &&
+ y_t^8 \dR (11/4 + 107\dR/12)
+ \ldots
,\\
c^{\lambda}_{3,3} &=& g_3^4 y_t^4 \dR C_F   \left (
-22 C_G/3 -24 C_F + 8 T_F N_q/3  
\right )
+ g_3^2 y_t^6 \dR C_F (15 + 18 \dR)
\nonumber \\ &&
+ y_t^8 \dR (-9/4 - \dR - 3 \dR^2)
+ \ldots
,\\
c^{y_t}_{1,1} &=& g_3^2 y_t (-3 C_F) + y_t^3  (\dR/2 + 3/4)
+ \ldots 
,
\\
c^{y_t}_{2,1} &=& g_3^4 y_t C_F \Bigl ( -\frac{97}{12} C_G - \frac{3}{4} C_F +
\frac{5}{3} T_F N_q \Bigr ) 
+ g_3^2 y_t^3 C_F \Bigl (3 + \frac{5}{4} \dR \Bigr )
+ y_t^5 \Bigl (\frac{3}{8} - \frac{9}{8} \dR \Bigr )  
+ \ldots
,
\phantom{xxx}
\\
c^{y_t}_{2,2} &=& g_3^4 y_t C_F \Bigl ( \frac{11}{2} C_G 
+ \frac{9}{2} C_F - 2 T_F N_q \Bigr )  
+ g_3^2 y_t^3  C_F \Bigl (-3 \dR - 9/2 \Bigr )
\nonumber \\ &&
+ y_t^5 \Bigl (\frac{3}{8}\dR^2 + \frac{9}{8} \dR + \frac{27}{32} \Bigr )  
+ \ldots
,\\
c^{g_3}_{1,1} &=& g_3^3 \Bigl (
-\frac{11}{6} C_G + \frac{2}{3} T_F N_q
\Bigr )
,
\\
c^{\phi}_{1,1} &=& -y_t^2 \dR 
+ \ldots
,\\
c^{\phi}_{2,1} &=& g_3^2 y_t^2 \left (-{5 \dR C_F}/{2} \right ) 
+ y_t^4 (9\dR/8) 
+ \ldots
,\\
c^{\phi}_{2,2} &=&  g_3^2 y_t^2 \left (3 \dR C_F \right ) 
+ y_t^4 (-3\dR/4) 
+ \ldots
.
\eeq
Here the ellipses refer to contributions that are known, but
are suppressed by couplings other than $y_t$ or $g_3$
to a sufficient extent that they are not pertinent for this paper.

Now, plugging eqs.~(\ref{eq:defphiB})-(\ref{eq:defzkB}) into the results
(\ref{eq:Veffbare}), (\ref{eq:V0bare}), (\ref{eq:V1bare}), (\ref{eq:V2bare})
and (\ref{eq:V3bare}) gives the effective potential in terms of renormalized
quantities. This can be written in a loop expansion as
\beq
V_{\rm eff} = \sum_{\ell= 0}^\infty \frac{1}{(16\pi^2)^\ell} V^{(\ell)}.
\eeq
Note that, as a convention, here the loop factors of $1/(16 \pi^2)^\ell$ have been extracted,
unlike the corresponding loop expansion in terms of bare parameters, eq.~(\ref{eq:Veffbare}).
In this section, I write
\beq
T &=& y_t^2 \phi^2/2,\\
H &=& m^2 + 3 \lambda \phi^2,\\
G &=& m^2 + \lambda \phi^2,
\eeq
for the $\MSbar$ field-dependent squared masses of the top quark, 
Higgs boson $H$,
and Goldstone bosons $G^{0}, G^\pm$ respectively, and define
\beq
\lnbar(X) \equiv \ln(X/Q^2)
\eeq
for $X = T,H,G$.
Retaining terms quadratic in $\lambda$ and $m^2$ in the one-loop part, 
and linear in $\lambda$ and $m^2$ in the two-loop part, and
taking the limit $\epsilon \rightarrow 0$, 
now gives
\beq
V^{(0)} &=& \Lambda + \frac{m^2}{2} \phi^2 + \frac{\lambda}{4} \phi^4 ,
\label{eq:V0ren}
\\
V^{(1)} &=& -\dR T^2 \bigl [\lnbar(T) - 3/2 \bigr ]
+ \frac{H^2}{4} \bigl [\lnbar(H) - 3/2 \bigr ]
+ \frac{3G^2}{4} \bigl [\lnbar(G) - 3/2 \bigr ]
,
\label{eq:V1ren}
\\
V^{(2)} &=& g_3^2 \dR C_F T^2 \bigl [6 \lnbar^2(T) - 16 \lnbar(T) + 18 \bigr ]
+ y_t^2 \dR T^2 \Bigl [-\frac{3}{2} \lnbar^2(T) + 8 \lnbar(T) - \frac{23}{2} - 
\frac{\pi^2}{6} \Bigr ]
\nonumber \\ &&
+ y_t^2 \dR T H \Bigl [
\frac{9}{2} + 4 \lnbar(T) - \frac{3}{2} \lnbar^2(T) +
\{ 1 - 3 \lnbar(T) \} \ln(H/T) 
\Bigr ]
\nonumber \\ &&
+ y_t^2 \dR T G \Bigl [
\frac{3}{2} + \frac{\pi^2}{3} + 2 \lnbar(T) - \frac{3}{2} \lnbar^2(T) +
3 \{ 1 - \lnbar(T) \} \ln(G/T) 
\Bigr ]
,\phantom{xxx}
\label{eq:V2ren}
\eeq
which agrees with the relevant limits of  
ref.~\cite{Ford:1992pn}, and the new result:
\beq
V^{(3)} &=& g_3^4 \dR C_F T^2 \Bigl \{
C_G \Bigr [-\frac{22}{3} \lnbar^3(T) 
             + \frac{185}{3} \lnbar^2(T)
             + (24 \zeta(3) - \frac{1111}{6}) \lnbar(T)
\nonumber \\ &&
             + \frac{2609}{12} 
             + \frac{44}{45}\pi^4 
             - \frac{232}{3} \zeta(3) 
             + \frac{16}{3} \ln^2(2) [\pi^2 - \ln^2(2)]
             - 128 {\rm Li}_4(1/2) 
\Bigl ]
\nonumber \\ &&
+
C_F \Bigr [-24 \lnbar^3(T) 
             + 63 \lnbar^2(T)
             - (48 \zeta(3) + \frac{121}{2}) \lnbar(T)
             + \frac{85}{12} 
             - \frac{88}{45}\pi^4 
\nonumber \\ &&
             + 192 \zeta(3) 
             - \frac{32}{3} \ln^2(2) [\pi^2 - \ln^2(2)]
             + 256 {\rm Li}_4(1/2) 
\Bigl ]
+
T_F \Bigr [48 \lnbar(T)
             - \frac{232}{3} 
             + 96 \zeta(3) 
\Bigl ]
\nonumber \\ &&
+
T_F N_q \Bigr [\frac{8}{3} \lnbar^3(T) 
             - \frac{52}{3} \lnbar^2(T)
              + \frac{142}{3} \lnbar(T)
             - \frac{161}{3} 
             - \frac{64}{3} \zeta(3) 
\Bigl ]
\Bigr \} 
\nonumber \\ &&
+ g_3^2 y_t^2 \dR C_F T^2
\Bigl \{
15 \lnbar^3(T) - 90 \lnbar^2(T) 
+ [
407/2
+ 3 \pi^2 + 
60 \zeta(3) 
] \lnbar(T)
- 54 \zeta(3)
\nonumber \\ &&
- \frac{2393}{12}
- \frac{29}{6} \pi^2 
+ \frac{31}{15} \pi^4 +
\frac{32}{3} \ln^2(2) [\pi^2 - \ln^2(2) ]
-256 {\rm Li}_4(1/2) 
\Bigr \}
\nonumber \\ &&
+ y_t^4 \dR T^2 \Bigl \{
-\frac{9}{4} \lnbar^3(T)
+\frac{57}{4} \lnbar^2(T)
+ \Bigl [-\frac{3}{4} \pi^2 
-\frac{121}{4} 
- 12 \zeta(3) \Bigr ] \lnbar(T)
\nonumber \\ &&
+ \frac{529}{24} 
+ \frac{23}{12}\pi^2 
- \frac{22}{45}\pi^4 
+ \frac{93}{2} \zeta(3) 
-\frac{8}{3} \ln^2(2) [\pi^2 - \ln^2(2)	]
+ 64 {\rm Li}_4(1/2) 
\Bigr \}
\nonumber \\ &&
+ y_t^4 \dR^2 T^2 \Bigl \{
\frac{7}{2} \lnbar^3(T)
+ [17/4 + 9 \ln(H/T) + 3 \ln(G/T)] \lnbar^2(T)
\nonumber \\ &&
+ \Bigl [
-\frac{659}{8}
-\frac{5}{6}\pi^2  
- 6 \ln(H/T) - 6 \ln(G/T) \Bigr ] \lnbar(T)
\nonumber \\ &&
+ \frac{4903}{48}
+ \frac{3}{4}\pi^2  
- 64 \zeta(3) 
+ \ln(H/T)
+ 3 \ln(G/T)
\Bigr \}.
\label{eq:V3ren}
\eeq
Note that
poles in $\epsilon$ are absent from eqs.~(\ref{eq:V1ren})-(\ref{eq:V3ren}); 
this is a non-trivial
check on the calculation, showing
agreement between the counter-term quantities $c^k_{\ell,n}$ and $c^\phi_{\ell,n}$ as extracted
from the known beta functions and anomalous dimension in the literature,
and as obtained from the diagrams calculated here.
This is equivalent to the check of renormalization group scale 
independence of the effective potential:
\beq
0 &=& Q\frac{\partial}{\partial Q} V^{(\ell)}
+ \sum_{\ell'=1}^\ell 
\Bigl [
\sum_k \beta^{(\ell')}_{k} \frac{\partial}{\partial z_k} 
- \gamma^{(\ell')} \phi\frac{\partial}{\partial \phi}
\Bigr ] V^{(\ell - \ell')},
\label{eq:RGVell}
\eeq
which follows from $dV_{\rm eff}/dQ = 0$. In fact, eq.~(\ref{eq:RGVell}) 
could have been used
to infer all of the terms in $V^{(3)}$ that contain $\lnbar(T)$,
just from knowledge of the 2-loop effective potential and the beta functions
and scalar anomalous dimension. 
I have checked this.

Plugging the Standard Model group theory constants 
of eqs.~(\ref{eq:defCG})-(\ref{eq:defNq})
into eq.~(\ref{eq:V3ren}) gives
\beq
V^{(3)} &=&
g_3^4 T^2 \Bigl \{
-184 \lnbar^3(T) 
+868 \lnbar^2(T)
+(32 \zeta(3) -\frac{5642}{3}) \lnbar(T)
             + \frac{16633}{9} 
             + \frac{176}{135}\pi^4 
\nonumber \\ &&
             + 32 \zeta(3)
             + \frac{64}{9} \ln^2(2) [\pi^2 - \ln^2(2)]
             - \frac{512}{3} {\rm Li}_4(1/2) 
\Bigr \}
\nonumber \\ &&
+g_3^2 y_t^2 T^2 \Bigl \{
60 \lnbar^3(T) - 360 \lnbar^2(T) + 
\bigl [814 + 12 \pi^2 + 240 \zeta(3) \bigr ] \lnbar (T)
-\frac{2393}{3}
\nonumber \\ &&
-\frac{58}{3} \pi^2
+ \frac{124}{15} \pi^4
+ \frac{128}{3}  \ln^2(2) [\pi^2 - \ln^2(2)]
- 216 \zeta(3) 
-1024 {\rm Li}_4(1/2)
\Bigr \}
\nonumber \\ &&
+ y_t^4 T^2 
\Bigl \{
\frac{99}{4} \lnbar^3(T) 
+ [81 + 81 \ln(H/T) + 27 \ln(G/T)] \lnbar^2(T)
\nonumber \\ &&
+ [-\frac{6657}{8} - \frac{39}{4}\pi^2 - 36 \zeta(3) 
- 54 \ln(H/T) -54 \ln(G/T)] \lnbar(T)
\nonumber \\ &&
+ \frac{15767}{16} + \frac{25}{2}\pi^2 - \frac{22}{15} \pi^4
- 8 \pi^2 \ln^2(2) + 8 \ln^4(2)
- \frac{873}{2} \zeta(3) 
\nonumber \\ &&
+ 192 {\rm Li}_4(1/2)
+ 9 \ln(H/T) + 27 \ln(G/T)
\Bigr \},
\label{eq:V3rensim}
\eeq
or, numerically,
\beq
V^{(3)} &\approx&
g_3^4 T^2 \Bigl \{
-184 \lnbar^3(T) 
+868 \lnbar^2(T)
-1842.2 
\,\lnbar(T)
+ 1957.3
\Bigr \} 
\nonumber \\ &&
+ g_3^2 y_t^2 T^2 \Bigl \{
60 \lnbar^3(T) - 360 \lnbar^2(T) + 1220.9 \lnbar(T) - 780.3
\Bigr \} 
\nonumber \\ &&
+ y_t^4 T^2 \Bigl \{
24.75 \lnbar^3(T) 
+ [81 + 81 \ln(H/T) + 27 \ln(G/T)] \lnbar^2(T)
\nonumber \\ &&
+ [-971.6 - 54 \ln(H/T) -54 \ln(G/T)]\lnbar(T)
+ 504.5 + 9 \ln(H/T) + 27 \ln(G/T)
\Bigr \}
.\phantom{xxxxx}
\label{eq:V3rennum}
\eeq 
Equation (\ref{eq:V3rensim}) or (\ref{eq:V3rennum}) may be consistently 
added to the full 2-loop effective potential
as given in ref.~\cite{Ford:1992pn}.


\section{The Goldstone boson catastrophe\label{sec:goldstone}}
\setcounter{equation}{0}
\setcounter{figure}{0}
\setcounter{table}{0}
\setcounter{footnote}{1}

Because of the doubled Goldstone boson
propagators
in diagrams (r) and (s) of Figure \ref{fig:diagrams3}, 
the three-loop effective potential has a logarithmic singularity in the 
limit $G = m^2 + \lambda \phi^2 = 0$, which corresponds to $\phi$ being at the minimum of the 
tree-level renormalized potential. In fact,  
the situation becomes progressively worse 
at higher loop orders, as these diagrams are part of a family
that also includes the one-loop Goldstone contributions and 
the two-loop diagrams
(b) and (c) in Figure \ref{fig:diagrams2}, and 
more generally, $\ell$-loop 
vacuum diagrams consisting of a ring of $\ell-1$
Goldstone boson propagators (all carrying the same momentum) 
punctuated by $\ell-1$ top (for $G^0$) 
or top/bottom (for $G^\pm$) one-loop
sub-diagrams. These diagrams give rise to 
contributions to the $\ell$-loop effective potential of the form\footnote{This sort of contribution 
has been noted before in refs.~\cite{Ford:1992mv,Einhorn:2007rv}
in the context of non-Goldstone scalars with small field-dependent
squared masses.}
\beq
V^{(\ell)} &\sim& (N_c y_t^2)^{\ell-1} T^2 (G/T)^{3-\ell} [\ln(G/T) + \ldots]. 
\label{eq:catastrophe}
\eeq
The ellipses in eq.~(\ref{eq:catastrophe}) includes constant terms.
At least for $\ell = 1,2,3$, higher powers of $\ln(G/T)$ are absent; from
eqs.~(\ref{eq:V1ren})-(\ref{eq:V3ren}) we see that at those 
loop orders one has specifically:
\beq
V^{(1)} &\sim& \frac{3}{4} G^2 \ln(G/T),
\label{eq:catone}
\\
V^{(2)} &\sim& -3 N_c y_t^2 T (\lnbar T-1) G \ln(G/T),
\label{eq:cattwo}
\\
V^{(3)} &\sim& 3 [N_c y_t^2 T (\lnbar T-1)]^2 \ln(G/T) .
\label{eq:catthree}
\eeq
Equation (\ref{eq:catastrophe}) means that at 4-loop order and higher, 
the singularity in $V_{\rm eff}$ 
as $G \rightarrow 0$ will be power-law, going like 
$1/G^{\ell-3}$ multiplied by terms constant and logarithmic in $G$.
Moreover, the first derivative of the effective potential 
with respect to $\phi$ diverges logarithmically in the $G=0$ limit even 
at two-loop order, and the second derivative already at one-loop order.

For a generic choice of renormalization scale, at the minimum of the 
full radiatively corrected effective potential, $G$ will be small 
(compared to $T$), but non-zero, and there is no true singularity. 
Nevertheless, the numerical effect can be non-trivial and can be quite 
important if one happens to choose a renormalization scale where $G$ is 
very close to 0.

The 
behavior of the effective potential for small $G$ that is illustrated in 
eqs.~(\ref{eq:catastrophe})-(\ref{eq:catthree}) seems quite troubling. Unlike 
similar situations where renormalization group improvement has been 
employed to study the behavior in the presence of small field-dependent 
masses in toy models, the fact that $G$ can be small in magnitude 
(and negative) 
is not associated with any 
real or apparent near-instability of the vacuum. Rather, small $G$ is just 
the expected and inevitable result for any spontaneously broken weakly 
gauged symmetry, even in a clearly stable vacuum. 
One might even have naively imagined that a 
particularly {\em good} choice of renormalization scale would be one that 
makes $G$ as small as possible (and positive), given that the Goldstone boson 
masses should be 0 when computed exactly 
(and the imaginary parts of the effective 
potential from negative $G$ do not really correspond to any instability 
in the theory). But, instead, a choice of 
renormalization scale that makes $G$ very small will actually provoke 
unphysically large contributions to the perturbatively computed 
effective potential and especially to its derivatives, and so apparently 
should be avoided.  It would be interesting to see in explicit detail 
how renormalization group improvement (or some other resummation 
or trick) can mitigate 
this behavior in the Standard Model case. 
However, I declare this to exceed 
the scope of the present paper.

\section{Numerical impact\label{sec:numerical}}
\setcounter{equation}{0}
\setcounter{figure}{0}
\setcounter{table}{0}
\setcounter{footnote}{1}

A full numerical study is also beyond the scope of this paper, but a few 
remarks about the practical impact of the results obtained above are in 
order. Consider, as a template, the central values of model parameters 
given in ref.~\cite{Buttazzo:2013uya}:
\beq
Q &=& M_t = \mbox{173.35 GeV}, 
\\
\lambda(M_t) &=& 0.12710,
\\
y_t(M_t) &=& 0.93697,
\\
g_3(M_t) &=& 1.1666,
\\
m^2(M_t) &=& -\mbox{(93.36 GeV})^2,
\\
g(M_t) &=& 0.6483,
\\
g'(M_t) &=& 0.3587.
\eeq
Now, minimizing the (real part of the) 
full two-loop effective potential of \cite{Ford:1992pn}, I obtain the 
Landau gauge $\MSbar$ VEV:
\beq
v(M_t)_{\mbox{2-loop}} &=& \mbox{247.25 GeV}.
\eeq
[At this minimum, one has $G = -(\mbox{30.76 GeV})^2$, 
so that the effective potential
computed in perturbation theory has an 
imaginary part due to $\ln(G)$ factors.]
If the three-loop contribution found above 
in eq.~(\ref{eq:V3rennum}) is included, 
I obtain instead
\beq
v(M_t)_{\mbox{3-loop}} &=& \mbox{246.91 GeV}
\eeq
for the same set of Lagrangian parameters. The majority of this shift 
comes from the $g_3^4$ contribution to $V^{(3)}$; if only those 
contributions were included, the VEV would be 246.84 GeV. 
However, beyond the 
observation that the effect of $V^{(3)}$ is to reduce the VEV
by about 0.34 GeV when all $\MSbar$ Lagrangian parameters are held fixed, 
this way 
of assessing the impact is of somewhat limited interest, because 
in the real world the Lagrangian parameters $m^2$ and $\lambda$ are not 
directly accessible.

Another exercise is to consider the relation between the 
physical Higgs mass $M_H$ and $\lambda$. Writing $V_{\rm eff} = V^{(0)} 
+ \Delta V$, the minimum of the potential $v \equiv \phi_{\rm min}$ is 
determined by $\partial V_{\rm eff}/\partial \phi = 0$, which allows us 
to eliminate $m^2$ according to
\beq
m^2 = -\lambda v^2 - \frac{1}{\phi} 
\frac{\partial(\Delta V)}{\partial \phi} \Bigl |_{\phi=v}
\eeq
The pole squared mass of the Higgs boson is 
determined from
\beq
M_H^2 = m^2 + 3 \lambda v^2 + \Pi_{HH}(M_H^2).
\eeq
where $\Pi_{HH}(s)$ is the self-energy function 
of the external momentum squared $s = -p^2$.
When evaluated at $s = 0$, $\Pi_{HH}$ 
coincides with the second derivative 
of the radiative part of the effective potential.
Thus we can write:
\beq
M_H^2 &=& m^2 + 3 \lambda v^2 + 
\frac{\partial^2 (\Delta V)}{\partial \phi^2} \Bigl |_{\phi = v}
+ \left [ \Pi_{HH}(M_H^2) - \Pi_{HH}(0)\right ]
\\
&=& 
2 \lambda v^2 + \left (\left [ -\frac{1}{\phi} \frac{\partial}{\partial \phi}
+ \frac{\partial^2}{\partial \phi^2}\right ] \Delta V \right )\Bigl |_{\phi = v}
+ \left [ \Pi_{HH}(M_H^2) - \Pi_{HH}(0)\right ].
\eeq
Now if we consider $M_H^2$ and $v$ as fixed inputs, and treat 
$\Pi_{HH}(M_H^2) - \Pi_{HH}(0)$ as small,
then we can estimate the change in $\lambda$ coming from 
inclusion of a new contribution to the effective potential $\delta V$ 
(e.g. 3-loop effects) as
\beq
\Delta \lambda &\approx& 
-\frac{1}{2 v^2} \Delta M_H^2 
\>\approx\> -\frac{1}{2 v^2} 
\left (\left [ -\frac{1}{\phi} \frac{\partial}{\partial \phi}
+ \frac{\partial^2}{\partial \phi^2}\right ] \delta V \right )\Bigl |_{\phi = v} .
\label{eq:deltalambda}
\eeq
The neglect of $\Pi_{HH}(M_H^2) - \Pi_{HH}(0)$ here is not entirely 
justified, even for diagrams that involve the top mass as the only other 
scale, because the expansion parameter $M_H^2/M_t^2 \approx 0.53$ is not 
so small. For some of the diagrams contributing to $\Pi_{HH}(s)$, 
the expansion 
in $s/M_t^2$ tends to have powers of the expansion variable with 
numerical coefficients that are smaller than 1, and the expansion in 
$M_H^2/M_t^2$ converges fairly quickly. However, terms of first order in 
$M_H^2/M_t^2$ can be quite significant. Furthermore, diagrams 
contributing to $\Pi_{HH}(M_H^2)$ in which the external momentum can be 
routed through the diagram in such a way as to miss all top-quark 
propagators will not be approximated well by $\Pi_{HH}(0)$ at all. 

In particular, this is true of some of the
diagrams involving the Goldstone bosons, 
notably the ones obtained from the vacuum diagrams described in the 
previous section by attaching two external $H^0$ legs. Those contributions
are not just wrong, but potentially very large.
The naive estimates from eq.~(\ref{eq:deltalambda}) 
and eqs.~(\ref{eq:catone})-(\ref{eq:catthree}) for the most singular
contribution as $G \rightarrow 0$ from each loop order $\ell = 1,2,3$ are:
\beq
\Delta M_H^2 \Bigl |_{\mbox{1-loop}} &\sim& 
\frac{6 \lambda^2 \phi^2}{16 \pi^2} \ln(G/T),
\\
\Delta M_H^2 \Bigl |_{\mbox{2-loop}} &\sim& 
-\frac{12 \lambda^2 \phi^2}{(16 \pi^2)^2}
\Bigl [ \frac{N_c y_t^2 T (\lnbar T-1)}{G} \Bigr ]
\\
\Delta M_H^2 \Bigl |_{\mbox{3-loop}} &\sim& 
-\frac{12 \lambda^2 \phi^2}{(16 \pi^2)^3}
\Bigl [ \frac{N_c y_t^2 T (\lnbar T-1)}{G} \Bigr ]^2 ,
\eeq
and for higher loop orders, using eq.~(\ref{eq:catastrophe}):
\beq
\Delta M_H^2 \Bigl |_{\ell\mbox{-loop}} &\sim& 
\frac{\lambda^2 \phi^2}{(16 \pi^2)^\ell} 
\Bigl [ \frac{N_c y_t^2 T}{G} \Bigr ]^{\ell-1},
\eeq
where the multiplicative numerical factors and logarithms are unknown.
These apparent singularities as $G \rightarrow 0$ 
are unphysical nonsense, and they cannot appear in the 
correct expression for $M_H^2$. 
The resolution is that they are canceled by
contributions to $\Pi_{HH}(M_H^2) - \Pi_{HH}(0)$, as one can check
explicitly at two-loop order. 

Since we do not yet have $\Pi_{HH}(s)$ at 3-loop order, we should 
certainly not attempt to estimate $\Delta \lambda$ (even roughly) using 
the part of $V^{(3)}$ involving $y_t^4$, because it
includes the offensive $\ln(G/T)$ [and $\ln(H/T)$] factors. However, we can still
make estimates of the contributions proportional to $g_3^4$ and $g_3^2 y_t^2$,
since at three-loop order these are not singular for $G \rightarrow 0$.
These should be taken only as estimates because, as mentioned above, corrections
from $[\Pi_{HH}(M_H^2) - \Pi_{HH}(0)]$
that go like $M_H^2/M_t^2$ can be significant, even when 
Goldstone boson shenanigans are absent. With this caveat, 
using eq.~(\ref{eq:deltalambda}), one obtains with the model parameters
listed above:
\beq
\Delta \lambda \Bigl |_{\mbox{3-loop}\> g_3^4\> \mbox{terms}} 
&=& -0.000014
\label{eq:deltalambdag34}
\\
\Delta \lambda \Bigl |_{\mbox{3-loop}\> g_3^2 y_t^2\> \mbox{terms}} 
&=& -0.000153
\label{eq:deltalambdag32}
\eeq
for a total of $\Delta \lambda = -0.000167$.
This can be compared to the theoretical error estimate used in 
ref.~\cite{Buttazzo:2013uya} of $\pm 0.00030$, and the parametric error
from the uncertainty on the Higgs mass of 
$0.000206 (\Delta M_H/\mbox{(100 MeV)})$.

It might seem somewhat surprising that the estimated shift in $\lambda$ from the 
$g_3^4$ contribution to $V^{(3)}$ is so much smaller than the $g_3^2 y_t^2$
effect, given that $g_3 > y_t$ and the numerical coefficients are larger
in the $g_3^4$ terms than in the $g_3^2 y_t^2$ terms. 
This is due to an accidental cancellation. 
To see how this works, consider a generic contribution to $V_{\rm eff}$ of the form:
\beq
\delta V = T^2 \left [a_0 + a_1 \lnbar(T) + a_2 \lnbar^2(T) + a_3 \lnbar^3(T) \right ].
\eeq
From eq.~(\ref{eq:deltalambda}), one obtains the estimate for the corresponding shift in $M_H^2$:
\beq
\Delta M_H^2 &=& 2 y_t^2 T \bigl [
(2 a_0 + 3 a_1 + 2 a_2) + 
(2 a_1 + 6 a_2 + 6 a_3) \lnbar(T) 
\nonumber \\ &&
+ (2 a_2 + 9 a_3) \lnbar^2(T) 
+ 2 a_3 \lnbar^3(T) \bigr ]
\eeq
Having chosen $Q = M_t$, the logarithms are small, $\lnbar(T) = -0.11315$, and so 
the largest contribution might, naively, 
be expected to come from the term that does not have $\lnbar(T)$ in it,
which is proportional to $2 a_0 + 3 a_1 + 2 a_2$.
However, for the one-loop contribution, 
\beq
(a_0,\, a_1,\, a_2,\, a_3)_{\mbox{1-loop}} = \frac{1}{16 \pi^2} (9/2,\, -3,\, 0,\, 0),
\eeq
so that $2 a_0 + 3 a_1 + 2 a_2$ happens to vanish.
At two loops, for the leading order in QCD:
\beq
(a_0,\, a_1,\, a_2,\, a_3)_{\mbox{2-loop}, g_3^2} = 
\frac{g_3^2}{(16 \pi^2)^2} (72,\, -64,\, 24,\, 0),
\eeq
and again $2 a_0 + 3 a_1 + 2 a_2$ happens to vanish. 
At three loops, for the leading order in 
QCD, 
\beq
(a_0,\, a_1,\, a_2,\, a_3)_{\mbox{3-loop}, g_3^4} = 
\frac{g_3^4}{(16 \pi^2)^3} (1957.3,\, -1842.2,\, 868,\, -184).
\eeq
Here the cancellation is not quite complete, but still
\beq
(2 a_0 + 3 a_1 + 2 a_2)_{\mbox{3-loop}, g_3^4} = 
\frac{g_3^4}{(16 \pi^2)^3} (124.1),
\eeq
which is well over an order of magnitude smaller than either $a_0$ or $a_1$ individually. 
Furthermore,
the $\lnbar(T)$ term has the opposite sign, and cancels about 40\% of this.  

In contrast, for the three-loop $g_3^2 y_t^2$ contribution, the individual coefficients are smaller,
\beq
(a_0,\, a_1,\, a_2,\, a_3)_{\mbox{3-loop}, g_3^2 y_t^2} = 
\frac{g_3^2 y_t^2}{(16 \pi^2)^3} (-780.3,\, 1220.9,\, -360,\, 60) ,
\eeq
but there is no efficient accidental cancellation in the term independent of $\lnbar(T)$:
\beq
(2 a_0 + 3 a_1 + 2 a_2)_{\mbox{3-loop}, g_3^2 y_t^2} = 
\frac{g_3^2 y_t^2}{(16 \pi^2)^3} (1382.2) .
\eeq

The preceding discussion points to an amusing fact. Suppose we took the 
``new" contribution to the effective potential $\delta V$ to consist of 
only the three-loop $g_3^4$ and $g_3^2 y_t^2$
contributions that do not include $\lnbar(T)$, on 
the grounds that the terms that do have $\lnbar(T)$ were all in 
principle known before this paper from the 2-loop effective potential 
and renormalization group invariance, by virtue of 
eq.~(\ref{eq:RGVell}). In other words, consider as the ``new" contribution:
\beq
\delta V = \frac{1}{(16 \pi^2)^3} T^2 \left [1957.3 g_3^4  - 780.3 g_3^2 y_t^2 \right ].
\eeq
From that point of view, we would find, instead of eqs.~(\ref{eq:deltalambdag34}) and
(\ref{eq:deltalambdag32}) above:
\beq
\Delta \lambda \Bigl |_{\mbox{3-loop}\> g_3^4\> \mbox{terms}} 
&=& -0.000710
\\
\Delta \lambda \Bigl |_{\mbox{3-loop}\> g_3^2 y_t^2\> \mbox{terms}} 
&=& 0.000182
\eeq
for a total of $\Delta\lambda = -0.000527$. The difference between this and the
value $\Delta\lambda = -0.000167$ obtained above is due to the subset of
$V^{(3)}$ terms dependent on $\lnbar(T)$.
Therefore, a well-meaning attempt to include
3-loop effects by using renormalization group invariance to obtain the $\lnbar(T)$ terms
in $V^{(3)}$ would have produced a spuriously large estimate for the
shift in $\lambda$, 
because it does not capture the accidental cancellations present
in the more complete calculation. In any case, the shift in $\lambda$ should really be calculated
using the full $M_H^2$ pole squared mass following from the three-loop $\Pi_{HH}(s)$. 
The effective potential found in this paper will allow a partial check of such a calculation through
comparison with the three loop 
$\Pi_{HH}(0) = \partial^2 (\Delta V)/\partial \phi^2 |_{\phi=v}$.

\section{Outlook\label{sec:outlook}}
\setcounter{equation}{0}
\setcounter{figure}{0}
\setcounter{table}{0}
\setcounter{footnote}{1}

The main new result of this paper is 
eq.~(\ref{eq:V3ren}), [or eq.~(\ref{eq:V3rensim}) or (\ref{eq:V3rennum})], which contains
the three-loop contributions to the effective potential in the Standard Model proportional to 
$m_t^4$ and to $g_3^4$, $g_3^2 y_t^2$, or $y_t^4$. In principle, this allows an improved
determination of the relation between the $\MSbar$ Lagrangian parameters and the VEV, 
although in practice one most deal with the fact that $m^2$ is not directly accessible.
The estimates of the numerical impact of the result, described in the previous section,
seem to suggest that the effects
are not large compared to the present parametric and other theoretical uncertainties,
although there is some accidental cancellation at work. 
While this is not unexpected, it is always a worthwhile goal to, if possible, 
reduce all theoretical errors far below the level where experimental
errors can compete with them, so that all 
uncertainties can be reliably blamed on experimentalists. 
Hopefully, the results above are one step in this direction.

\section*{Appendix: Individual diagram contributions}\label{sec:appendix}
\renewcommand{\theequation}{A.\arabic{equation}}
\setcounter{equation}{0}
\setcounter{footnote}{1}

The individual contributions  
to eq.~(\ref{eq:V3bare}) from the diagrams in Figure \ref{fig:diagrams3} are:
\beq
V^{(3,a)} &=& g_{3B}^4 \dR C_F (C_F - C_G/2) 
\biggl \{
\frac{(7-2d)(d-5) (d-2)^3}{(d-4)^2 (d-3)^2} \masterJ 
\nonumber \\ &&
+ \Bigl [
\frac{2000 - 3656 d + 2643 d^2  - 939 d^3  + 163 d^4  - 11 d^5}{2 (d-4)(d-3)(2d-7)}
+ \frac{(d-3)(3d^2 - 11 d + 8)}{2d-7} \xi 
\nonumber \\ &&
+ \frac{(d-4)(d-3)(d-1)}{2(2d-7)} \xi^2
\Bigr ] \masterK
+ \frac{(d-6)(d-3)(d^2 - 7d + 8)}{2 (d-4)^2} \masterL
\biggr \}
,
\\
V^{(3,b)} &=& g_{3B}^4 \dR C_F^2 \biggl \{
\frac{(d-2)^2 (d-1)^2}{2 (3-d)} \masterJ 
+
\Bigl [
\frac{(d-1)^2 (3d-8)}{2(2d-7)} 
+ \frac{(1-d)(3d^2 - 17d + 24)}{2d-7} \xi
\nonumber \\ &&
+ \frac{(1-d)(d^2 - 7d + 12)}{2 (2d-7)} \xi^2
\Bigr ] \masterK
\biggr \}
,
\\
V^{(3,c)} &=& g_{3B}^4 \dR C_F C_G \biggl \{
\frac{(2-d)^3}{(d-3)^2} \masterJ
+ 
\Bigl [
\frac{7d^4 - 67 d^3 + 237 d^2 - 373 d + 226}{2(3-d)(2d-7)} 
\nonumber \\ &&
+ \frac{3(d-1)(d-3)^2}{2d-7} \xi
+ \frac{(d-4)(d-3)(d-1)}{2(2d-7)} \xi^2
\Bigr ]
\masterK
\biggr \}
,
\\
V^{(3,d)} &=& g_{3B}^4 \dR C_F C_G \frac{d-3}{4 (2d-7)}\Bigl [ 
7 d^2 - 19 d + 14 + 2(1-d)(3d-10) \xi 
\nonumber \\ &&
+ (4-d)(d-1) \xi^2
\Bigr ] \masterK
,
\\
V^{(3,e)} &=& g_{3B}^4 \dR C_F C_G \frac{d-3}{2 (2d-7)} \masterK
,
\\
V^{(3,f)} &=& g_{3B}^4 \dR C_F T_F \biggl [
\frac{2(5-d)(d-2)^3}{(d-6)(d-4)(d-3)} \masterJ
+ \frac{d^3 - 7 d^2 + 6 d + 16}{(d-4)(6-d)} \masterL
\biggr ]
,
\\
V^{(3,g)} &=& g_{3B}^4 \dR C_F T_F (N_q-1) \frac{4(3-d)(d-2)}{2d-7} \masterK
,
\\
V^{(3,h)} &=& g_{3B}^2 y_{tB}^2 \dR C_F \biggl \{
\frac{(2-d)(d^4 - 8 d^3 + 17 d^2 + 8 d - 44)}{(d-4)^2 (d-3)^2}\masterJ
+ \frac{(3-d)(d^2 + 2d - 16)}{2 (d-4)^2} \masterL
\nonumber \\ && 
+
\Bigl [
\frac{5 d^4 - 60 d^3 + 283 d^2 - 618 d + 520}{(4-d)(d-3)(2d-7)}
+
\frac{(d-6)(d-3)}{2d-7} \xi
\Bigr ] \masterK
\biggr \}
,
\\
V^{(3,i)} &=& g_{3B}^2 y_{tB}^2 \dR C_F 
\frac{d-3}{d-4} \left \{
4 (d-3) \masterK
+ (4-2d) \masterQ
+ (2d-4) \Bigl [\frac{d}{d-4} - 2\xi \Bigr ] \masterP
\right \}
,
\\
V^{(3,j)} &=& g_{3B}^2 y_{tB}^2 \dR C_F \Bigl \{
\frac{(d-2)(d-1)}{d-3} \masterJ
+ \Bigl [
\frac{(1-d)(3d-8)}{2d-7} + \frac{(6-d)(d-3)}{2d-7}\xi 
\Bigr ] \masterK
\Bigr \}
,
\\
V^{(3,k)} &=& g_{3B}^2 y_{tB}^2 \dR C_F \Bigl \{
\Bigl [
\frac{(2-d)(d-1)(3d-8)}{2 (d-4)(d-3)} + \frac{2 (d-3)(d-2)}{d-4}\xi
\Bigr ] \masterP
\nonumber \\ &&
+
\frac{(2-d)(d-1)(2d-5)}{2(d-3)} \masterQ \Bigr \}
,
\\
V^{(3,l)} &=& g_{3B}^2 y_{tB}^2 \dR C_F \frac{2(d-3)(d-2)}{d-4} \xi \masterP
,
\\
V^{(3,m)} &=& y_{tB}^4 \dR  \Bigl [
-\frac{(d-2)^2 (d^2 -11 d + 26)}{4 (d-4)^2 (d-3)^2}\masterJ
+ \frac{(3d-8)(d^2-4d+2)}{2(d-4)(d-3)(2d-7)} \masterK
\nonumber \\ &&
+ \frac{(6-d)(3d-8)}{8 (d-4)^2} \masterL
\Bigr ]
,
\\
V^{(3,n)} &=& y_{tB}^4 \dR  \Bigl [
\frac{1}{2(3-d)}\masterJ +
\frac{d^2-12d+26}{2(2-d)(2d-7)} \masterK
\Bigr ]
,
\\
V^{(3,o)} &=& y_{tB}^4 \dR  \Bigl [
\frac{4 d^2 - 21 d + 28}{2 (d-4)(d-3)} \masterP 
+ \frac{2d-5}{2(d-3)}\masterQ \Bigr ]
,
\\
V^{(3,p)} &=& y_{tB}^4 \dR  \frac{3(2-d)}{8} \masterR
,
\\
V^{(3,q)} &=& y_{tB}^4 \dR \Bigl [
\frac{d^2 - 10 d + 20}{d(d-2)}\masterK + 
\frac{2}{d} \masterQ \bigr ]
,
\\
V^{(3,r)} &=& y_{tB}^4 \dR^2 \Bigl \{
\Bigl [\frac{2(d-2)}{(d-4)(d-3)} 
+ \frac{d-2}{2} (x_H/x)^{d/2-2}
+ \frac{d-2}{2} (x_G/x)^{d/2-2}
\Bigr ]\masterJ 
\nonumber \\ &&
+ \frac{d-2}{d-4} \masterL 
+ 
4 x \frac{d}{dx_H} \Bigl [ 
x I_{xxxxx_H}
- A I_{xxx_H} 
\Bigr ]
\Bigr \}
,
\\
V^{(3,s)} &=& y_{tB}^4 \dR^2 \Bigl \{
\frac{d-2}{4} (x_G/x)^{d/2-2} J + 
\frac{18-7d}{2(d-2)} \masterK + 3 \masterQ 
+ x \frac{d}{dx_G} \Bigl [
\frac{x}{2} I_{xx00x_G}
- A I_{x0x_G} 
\Bigr ]
\Bigr \}
.
\phantom{xxxx.}
\eeq
The sums of these contributions gives eq.~(\ref{eq:V3bare}). 
The cancellation of the dependence on the QCD gauge-fixing parameter $\xi$ provides a useful check.

{\it Acknowledgments:} 
This work was supported in part by the National Science Foundation grant number PHY-1068369. 
This research was supported in part by the National Science Foundation under Grant No. NSF PHY11-25915.


\end{document}